\documentclass[twocolumn]{aastex63}
\usepackage{amsmath}
\usepackage{ulem}

\shorttitle{WARP}
\shortauthors{Hamano et al.}

\begin{document}

\title{WARP: The Data Reduction Pipeline for the WINERED spectrograph}

\correspondingauthor{Satoshi Hamano}

\author[0000-0002-6505-3395]{Satoshi Hamano}
\affiliation{National Astronomical Observatory of Japan, 2-21-1 Osawa, Mitaka, Tokyo 181--8588, Japan}
\affiliation{Laboratory of Infrared High-resolution spectroscopy(LiH), Koyama Astronomical Observatory, Kyoto Sangyo University, Motoyama, Kamigamo, Kita-ku, Kyoto 603--8555, Japan}
\email{satoshi.hamano@nao.ac.jp}
\author[0000-0003-2380-8582]{Yuji Ikeda}
\affiliation{Laboratory of Infrared High-resolution spectroscopy(LiH), Koyama Astronomical Observatory, Kyoto Sangyo University, Motoyama, Kamigamo, Kita-ku, Kyoto 603--8555, Japan}
\affiliation{Photocoding, 460-102 Iwakura-Nakamachi, Sakyo-ku, Kyoto, 606--0025, Japan}
\author{Shogo Otsubo}
\affiliation{Laboratory of Infrared High-resolution spectroscopy(LiH), Koyama Astronomical Observatory, Kyoto Sangyo University, Motoyama, Kamigamo, Kita-ku, Kyoto 603--8555, Japan}
\author{Haruki Katoh}
\affiliation{Laboratory of Infrared High-resolution spectroscopy(LiH), Koyama Astronomical Observatory, Kyoto Sangyo University, Motoyama, Kamigamo, Kita-ku, Kyoto 603--8555, Japan}
\affiliation{Department of Physics, Faculty of Sciences, Kyoto Sangyo University, Motoyama, Kamigamo, Kita-ku, Kyoto 603--8555, Japan}
\author{Kei Fukue}
\affiliation{Laboratory of Infrared High-resolution spectroscopy(LiH), Koyama Astronomical Observatory, Kyoto Sangyo University, Motoyama, Kamigamo, Kita-ku, Kyoto 603--8555, Japan}
\affiliation{Education Center for Medicine and Nursing, Shiga University of Medical Science, Seta Tsukinowa-cho, Otsu, 520-2192 Shiga, Japan}
\author{Noriyuki Matsunaga}
\affiliation{Department of Astronomy, Graduate School of Science, University of Tokyo, Bunkyo-ku, Tokyo 113--0033, Japan}
\affiliation{Laboratory of Infrared High-resolution spectroscopy(LiH), Koyama Astronomical Observatory, Kyoto Sangyo University, Motoyama, Kamigamo, Kita-ku, Kyoto 603--8555, Japan}
\author[0000-0002-2861-4069]{Daisuke Taniguchi}
\affiliation{National Astronomical Observatory of Japan, 2-21-1 Osawa, Mitaka, Tokyo 181--8588, Japan}
\author[0000-0003-2011-9159]{Hideyo Kawakita}
\affiliation{Laboratory of Infrared High-resolution spectroscopy(LiH), Koyama Astronomical Observatory, Kyoto Sangyo University, Motoyama, Kamigamo, Kita-ku, Kyoto 603--8555, Japan}
\affiliation{Department of Physics, Faculty of Sciences, Kyoto Sangyo University, Motoyama, Kamigamo, Kita-ku, Kyoto 603--8555, Japan}
\author{Keiichi Takenaka}
\affiliation{Laboratory of Infrared High-resolution spectroscopy(LiH), Koyama Astronomical Observatory, Kyoto Sangyo University, Motoyama, Kamigamo, Kita-ku, Kyoto 603--8555, Japan}
\affiliation{Department of Physics, Faculty of Sciences, Kyoto Sangyo University, Motoyama, Kamigamo, Kita-ku, Kyoto 603--8555, Japan}
\author{Sohei Kondo}
\affiliation{Kiso Observatory, Institute of Astronomy, School of Science,The University of Tokyo, 10762-30 Mitake, Kiso-machi, Kiso-gun, Nagano, 397--0101, Japan}
\affiliation{Laboratory of Infrared High-resolution spectroscopy(LiH), Koyama Astronomical Observatory, Kyoto Sangyo University, Motoyama, Kamigamo, Kita-ku, Kyoto 603--8555, Japan}
\author[0000-0001-6401-723X]{Hiroaki Sameshima}
\affiliation{Institute of Astronomy, School of Science, University of Tokyo, 2-21-1 Osawa, Mitaka, Tokyo 181--0015, Japan}
\affiliation{Laboratory of Infrared High-resolution spectroscopy(LiH), Koyama Astronomical Observatory, Kyoto Sangyo University, Motoyama, Kamigamo, Kita-ku, Kyoto 603--8555, Japan}

\begin{abstract}

We present a data reduction pipeline written in Python for data obtained with the near-infrared cross-dispersed echelle spectrograph, WINERED, which yields a 0.91$-$1.35 $\mu$m spectrum with the resolving power of $R_{\text{max}} \equiv \lambda / \Delta \lambda = 28,000$ or 70,000 depending on the observing mode. The pipeline was developed to efficiently extract the spectrum from the raw data with high quality. It comprises two modes: the calibration and the science mode. The calibration mode automatically produces the flat-fielding image, bad pixel map, echellogram distortion map and the dispersion solution from the set of the calibration data. Using calibration images and parameters, the science data of astronomical objects can be reduced automatically using the science mode. The science mode is also used for the real-time quick look at the data during observations. An example of the spectra reduced with WARP is presented. The effect of the highly inclined slit image on the spectral resolution is discussed.
\end{abstract}


\section{Introduction}

The amount of data produced by astronomical observations is rapidly increasing. Accordingly, the pipeline software, which takes into account instrument characteristics, holds increasing importance for the efficient production of quality-controlled scientific data. The cross-dispersed echelle spectrograph is an instrument, for which the pipeline plays an important role, as the data reduction procedures for the echelle data are complicated due to the tilt of the dispersion and spatial directions relative to the detector coordinates \citep{pis02,pis21}. Furthermore, numerous research projects using the high-resolution echelle spectrographs demand high accuracies of the wavelengths and fluxes of the obtained spectra \citep[e.g.,][]{dor23}. Therefore, the pipeline for the echelle spectrograph must satisfy these strict requirements.

Recently, several cross-dispersed near-infrared (NIR) spectrographs whose spectral resolutions and/or wavelength coverages are comparable to those of optical high-resolution spectrographs have been developed \citep{kae04,ray12,qui13,fol14,ori14,cla18,ray22}. Compared to the data obtained from optical spectrographs, some difficulties are present in the NIR spectroscopic data, namely, strong night sky emission and telluric absorption lines. Furthermore, the problematic persistence, the afterimage of earlier exposures, appears due to the characteristics of IR arrays \citep{tul19}. Therefore, consideration of these systematic uncertainty is necessary for the pipeline software for the NIR spectroscopic data.

In this study, we present a pipeline software developed for the WINERED spectrograph \citep{ike22}, a cross-dispersed NIR spectrograph spanning the 0.91--1.35 $\mu$m range with high resolution ($R\equiv \lambda / \Delta \lambda = 28000$ and 70000). We call the pipeline software ``WARP'' (WINERED Automatic Reduction Pipeline). WINERED uses a 1.7 $\mu$m cutoff 2048 $\times$ 2048 HAWAII-2RG infrared array. Three observational modes are equipped with the WINERED: the WIDE mode (0.91--1.35 $\mu$m with $R = 28000$), the HIRES-Y mode (0.96--1.11 $\mu$m with $R = 70000$), and the HIRES-J mode (1.13--1.35 $\mu$m with $R = 70000$). One of the difficulties in the reduction of the WINERED data is the large tilt of the slit image on the detector. Because WINERED adopted non-white-pupil system to maximize the throughput of the optics, the $\gamma$ angle of the echelle grating must be large to avoid the mechanical interference between the collimator and the cross-disperser, resulting in the large tilt on the detector \citep[see Section 3.1 of][]{ike22}. The angles of this spatial direction from the axis of the array coordinate are about 33 and 50 degrees for WIDE and HIRES modes, respectively. The tilt is particularly large for HIRES modes because of the larger $\gamma$ angle. The large distortion of the slit image on the array complicates the reduction processes and degrades of the spectral resolution from the expected spec. These instrumental characteristics are taken into consideration in the development of WARP. By investigating such characteristics of the WINERED data and by developing of the reduction procedure protocols, we succeeded to automatize almost all reduction procedures of the WINERED data in WARP. The rest of this paper is organized as follows. The overview of the WINERED instrument, its observations, and the calibration data is described in Section 2. The overall design of the pipeline is presented in Section 3. In Section 4 and 5, the reduction flow and protocols of the calibration and science modes are presented, respectively.

\section{WINERED}
\subsection{Instrument overview}

WINERED, the NIR high-resolution spectrograph, is a PI-type instrument commissioned in June 2012. WINERED has three observational modes: the WIDE mode (0.91--1.35 $\mu$m with $R_{\text{max}}\equiv28000$), the HIRES-Y mode (0.96--1.11 $\mu$m with $R_{\text{max}}\equiv70000$), and the HIRES-J mode (1.13--1.35 $\mu$m with $R_{\text{max}}\equiv70000$). The WIDE mode employs a classical echelle grating while the HIRES modes employ a high-blazed echelle grating. The optical paths between the WIDE and two HIRES modes can be switched by inserting or ejecting the Au fold mirror. Two HIRES modes use different VPH cross-dispersers, which are designed to be switchable by the remote control. The WIDE, HIRES-Y, and HIRES-J modes use the echelle orders of $m=42$--61, 159--184, and 131--155, respectively. 

It was operated at the Nasmyth focus of the Araki 1.3-m telescope at Koyama Astronomical Observatory, Kyoto Sangyo University until the end of the 2016. In January 2017, WINERED moved to the La Silla observatory and was mounted on the New Technology Telescope ($D=3.58$ m). Since July 2022, WINERED has been mounted on Magellan telescope ($D=6.5$ m) at the Las Campanas observatory. One of the characteristics of WINERED is the unprecedentedly high throughputs: over 50\% and 40\% for WIDE and HIRES modes, respectively, including the array Q.E.. The high throughputs and low noise of the WINERED spectrograph enable high-quality (high signal-to-noise ratio) spectra comparable to that of optical spectrographs. Therefore, the pipeline software must maximize the quality of WINERED spectra. The technical details of the WINERED spectrograph are fully described in \citet{ike22}.

Three slits with the widths of 2, 2.8, and 4 pixels on the array are installed in the WINERED. The length of all slits amounts to 60.5 pixels. Multi-pinholes are likewise installed for the calibration data to characterize aperture positions and curvatures. Table \ref{slit} shows the spectral resolving powers and dimensions for each slit and telescope. 

\begin{deluxetable*}{ccccccccccccc}
\tabletypesize{\scriptsize}
\tablecaption{Spectral resolving powers and slit dimensions \label{slit}}
\tablehead{
 \colhead{} & \colhead{} & \colhead{} & \multicolumn{2}{c}{$R$} & \colhead{}& \multicolumn{3}{c}{Width (arcsec)} & \colhead{}& \multicolumn{3}{c}{Length (arcsec)} \\  \cline{4-5} \cline{7-9} \cline{11-13}
 \colhead{Slit} & \colhead{Width (pix)} & \colhead{Length (pix)} & \colhead{WIDE} & \colhead{HIRES} & \colhead{}& \colhead{Araki} & \colhead{NTT}  & \colhead{Magellan}  & \colhead{}& \colhead{Araki} & \colhead{NTT}  & \colhead{Magellan}
}
\startdata
100 $\mu$m slit & 2 &  &  28000 & 70000 &  & 1.5 & 0.54 & 0.3 &  & & & \\
140 $\mu$m slit & 2.8 & 60.5 &  20000 & 50000 &  & 2.1 & 0.76 & 0.42 &  & 45 & 16.34 & 9.0 \\
200 $\mu$m slit & 4 & & 14000 & 35000 &  & 3.0 & 1.08 & 0.6 &  & & &
\enddata
\end{deluxetable*}

\begin{figure*}
\includegraphics[width=18cm,clip]{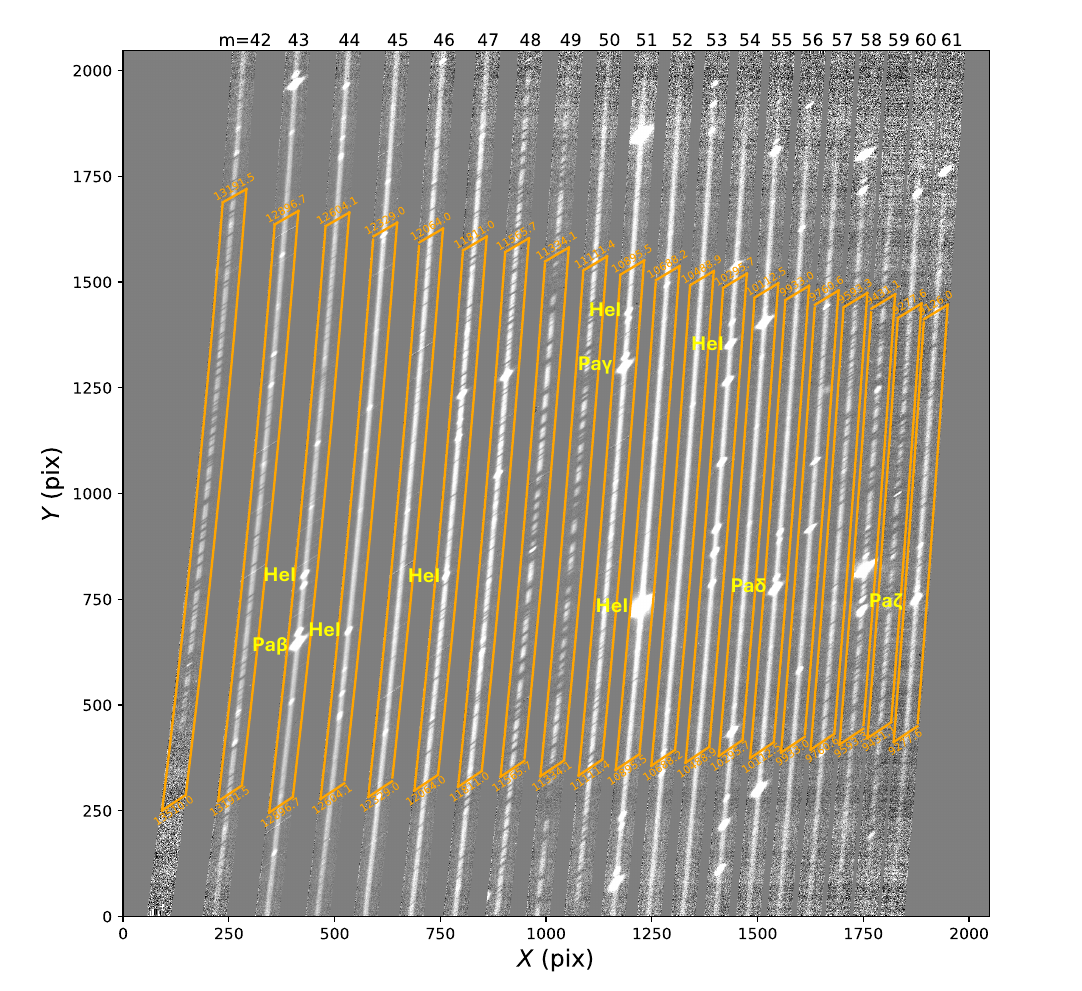}
\caption{Example of a scientific frame after flat fielding obtained with the WINERED WIDE mode (0.91--1.35 $\mu$m, $R$=28000). A very bright planetary nebula, NGC 7027, is observed with an exposure time of 600 sec. The area of the free spectral range is overplotted. Some prominent emission lines of H and He are marked with their labels. The absorption lines seen in the continuum are mostly caused by strong telluric absorption lines. }
\label{2dimage}
\end{figure*}

Figure \ref{2dimage} shows the echellogram of WINERED data. The size of the frame is 2048 $\times$ 2048 pixels. Reference pixels are arranged along the frame of the array with a width of 2 pixels. We define $X$ and $Y$ as the array coordinates as shown in the figures illustrated throughout this study. The $X$ axis is the dispersion direction of VPH cross-dispersers while the $Y$ axis is the dispersion direction of the echelle grating. As the echelle order $m$ increases, the X coordinate of each order becomes smaller, and the length of the spectrum increases because the free spectral range is equal to $\lambda_c / m$ ($\lambda_c$: the central wavelength of each order). Accordingly, the longer the wavelength of the spectrum, the smaller its $Y$ coordinates. One of the characteristics of WINERED data is the large tilted slit images with respect to the detector coordinates. The shape of slit images on the detector can be assumed as a straight line. The tilt angle depends strongly on the $Y$ coordinates and weakly on the echelle order $m$. The detailed characteristics of the tilt angle and the resampling method correcting the tilt are described in Section 4.4.

Because all optical components of WINERED, except for the camera optics and the array, are maintained at the room temperature, the spectra positions on the array drift slightly in response to the variation of the ambient temperature. The drift of the slit images during observations can degrade the spectral resolution and wavelength accuracy. The typical wavelength uncertainty due to the wavelength shift was estimated as $\Delta (d \lambda/ \lambda) <$ 1/10 pixel, which corresponds to the typical variation of the air temperature at the NTT site with $\Delta T \sim 5$K \citep{ike22}. We correct the relative wavelength shifts between the frames in a dataset by correlating the spectra extracted from the frames (Section 5.7). 

The infrared array H2RG shows the latent image after the illumination of the array, which is referred to as ``persistence''. The persistence image has a poor influence both on the  observation strategy and the pipeline reduction.
It appears by the emission of electrons trapped by pixel defects during the previous illumination. The brighter the illumination, the stronger the persistence image in the following frames gets. The persistence effect is difficult to avoid by nodding the telescope due to the narrow slit length. 
Due to the complicated behavior of the persistence, we failed to make a predictive persistence model for the WINERED array to remove the persistence signal from the images.
The WINERED H2RG shows about 1\% of the previous signal just after resetting. After the observation of bright objects (e.g., telluric standard stars), the same slit position should not be used for observations of faint targets for several hours. We considered developing the pipeline such that the persistence signal does not blend with the target spectrum. 

The high throughput and resolution of WINERED have advanced the observational studies of various astronomical objects: metallicities and chemical abundances of various type of stars \citep{dor18,kon19,jia20,fuk21,tan21,mat22,mat23}, the diffuse interstellar bands and interstellar molecules \citep{ham19,ham22}, young stellar objects \citep{yas19}, AGN \citep{miz23}, luminous blue variable stars \citep{miz18}, quasars\citep{sam20} and comets \citep{shi17}.

\subsection{Observation procedures}
In the ground-based NIR observations, we obtain spectroscopic data by nodding the telescope to cancel the sky emission. We adopt two types of nodding patterns: ``ABBA'' (nodding along the slit without a sky exposure) and ``OSO'' (nodding between the object and the sky region). The extended and/or crowded objects are observed with the OSO sequence. The bright objects are also observed with the OSO sequence to reduce the area affected by the persistence signal. In the other cases, the objects are observed with the ABBA sequence, which is more efficient. The position of the target in a slit is recorded in the header of the fits file and the observation database (WINERED Observational Database System: WODB). The pipeline refers to the position information during the data reduction processes for determining target positions.

Telluric standard stars should be observed to cancel the telluric absorption lines in the target spectra. We prepared the list of the bright A0V stars. During observations, we select telluric standard stars whose airmasses are similar to those of the targets. The configuration of the observations, such as the slit and the observing mode, must be the same. Since the telluric absorption lines are variable with time, we obtain the spectrum of the telluric standard star once in a few hours even if the airmass of the targets is similar over a long time. The correction of the telluric absorption lines using the A0V star spectrum is provided in a separate software, because the removal of the stellar lines from the telluric star's spectrum needs for the complicated procedures. Our method of the telluric correction is presented in \citet{sam18}.

\subsection{Calibration}
Calibration data are usually obtained just after the observations. We do not obtain them during or just before the observations to avoid the influence of the persistence signal. For flat fielding, the flat on and off images are obtained by observing the dome flat with and without the illumination with a lamp. 
For wavelength calibration, we use a ThAr lamp, which is installed in the calibration unit. This lamp illuminates an entire slit at F/11 through the relay optics. Diffuser-plates are placed at both the object and pupil positions of the relay optics, achieving sufficiently high spatial uniformity \citep[see Figure 8 in][]{ike22}.
The flat lamp image with multi-pinholes mounted in the slit stocker is likewise necessary for the aperture map. To minimize the influence of the persistence signal on flat-fielding images, the calibration data are obtained in the following order: (1) off-flat-fielding images of the slits that will be or were used in observations, (2) on-flat-fielding images of the same slits, (3) on-flat-fielding images of multi-pinholes, and (4) comparison ThAr lamp images of the slits. The flat and comparison images must be acquired with the same instrument configurations as those used in the actual observations (slits and observing modes). Multi-pinholes images must likewise be acquired with the same observing modes as those used in the observations.

\section{Pipeline overview}

WARP\footnote{https://github.com/SatoshiHamano/WARP. The user documentation is provided as well as the software via the github page.} aims to realize a fully automatic reduction from raw data to the final products. WARP comprises two modes: a science and a calibration mode. The science mode reduces the WINERED raw data of astronomical objects to the final wavelength-calibrated spectra without any manual operation. The set of processed calibration data and parameters is necessary for the use of science mode. The WARP calibration mode can automatically produce the set of calibration files from the combined images of flat and comparison lamps. The details on data processing in the calibration and science modes are given in Section 4 and 5, respectively.

WARP is written in {\tt Python}, and uses the packages of {\tt PyRAF\footnote{A command language for running {\tt IRAF} based on the {\tt Python} scripting language. {\tt PyRAF} was developed by NOAO. The official support and development of PyRAF were ended recently.}} and {\tt astropy}\footnote{A community-developed core {\tt Python} package for astronomy \citep{astropy:2013, astropy:2018}.} for handling astronomical data. WARP also uses {\tt numpy}, {\tt scipy}, and {\tt matplotlib} for scientific calculation and data visualization.

WARP can be used as a command-line tool. Although it does not offer a graphic user interface, the results of the data processing can be verified with various figures and the reduction report pdf file made at the last step of the WARP processing. 

\section{Calibration mode}

\begin{figure*}
\includegraphics[width=9cm,clip]{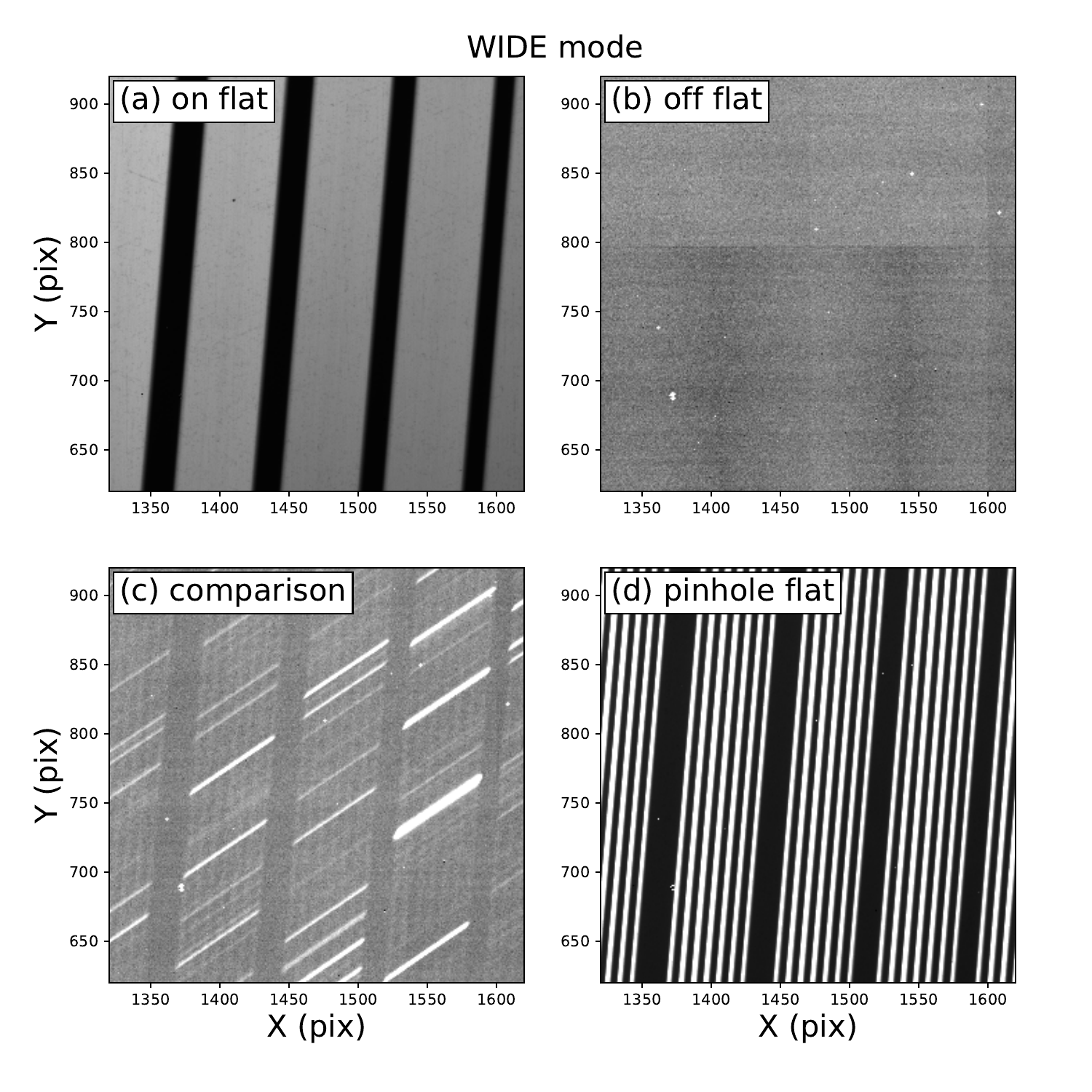}
\includegraphics[width=9cm,clip]{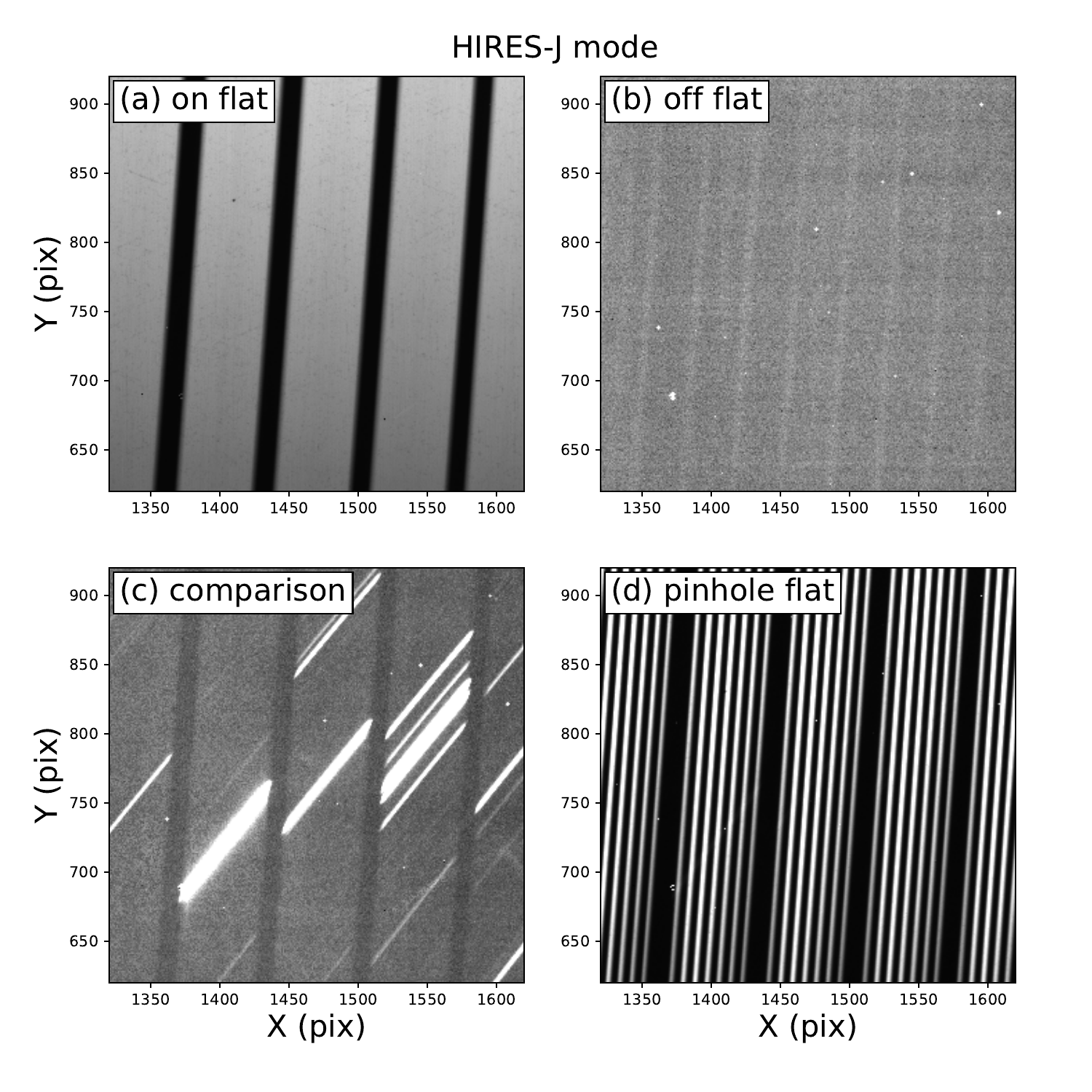}
\caption{Combined images of on-flat (a), off-flat (b), comparison (c), and pinhole flat data (d) obtained with WIDE (left panel) and HIRES-J (right panel) modes.}
\label{calib}
\end{figure*}

The WARP calibration mode reduces the calibration raw data and produces the dataset of calibration images and parameters required for the reduction of raw spectrum data with the WARP science mode.
To run the calibration mode, the necessary input includes the raw data of the on-flat-fielding image, off-flat-fielding image, multi-pinholes flat image, and the comparison image of ThAr lamps (Figure \ref{calib}). Usually, the user prepare the combined images of multiple raw frames for increasing the S/N ratios. Furthermore, the ascii file, which lists the keywords (``onflat'', ``offflat'', ``pinhole'', and ``comp'') and corresponding fits files, is necessary as input to give the WARP the file names of the prepared input data and their data types. 

\begin{figure*}
\includegraphics[width=16cm,clip]{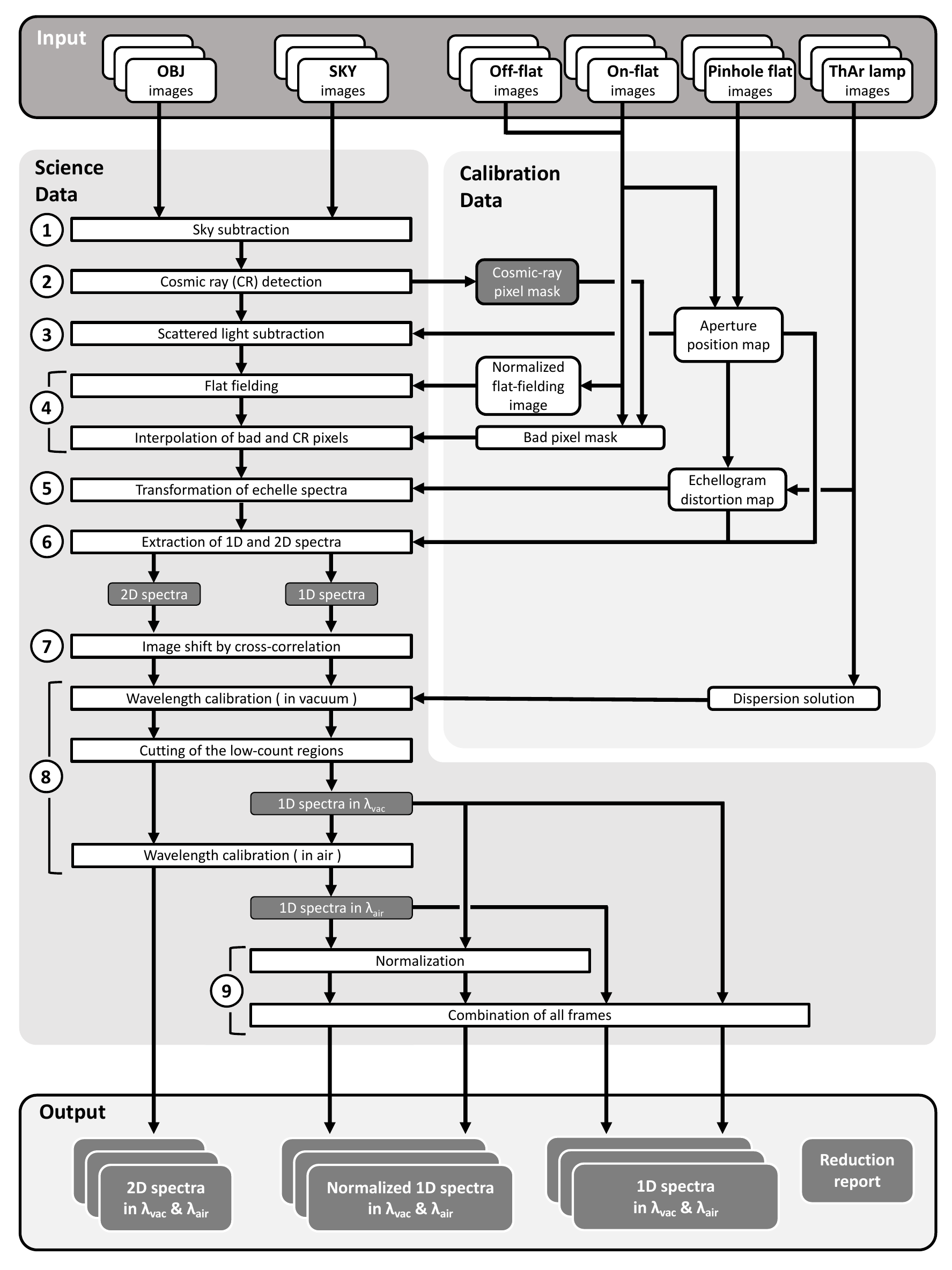}
\caption{WARP reduction procedure flow.}
\label{flow}
\end{figure*}

With the calibration mode, following output can be obtained automatically. 
\begin{itemize}
\item Echelle order map: information on the position and shape of the apertures for each echelle order in the raw data measured from the multi-pinholes flat data.
\item Bad pixel mask: a map of the pixels that lose their sensitivities (dead pixels), and pixels with very high dark current (hot pixels). The mask is made from the on and off flat-fielding data.
\item Echellogram distortion map: a distortion map of the slit images for each echelle order calculated from the aperture position map and the tilts of ThAr emission lines in the comparison data.
\item Dispersion solution: the wavelength function of the pixels in the dispersion direction and the echelle orders made from the ThAr emission lines in the spectra of the comparison data.
\end{itemize}
With this dataset, astronomical spectra can be produced with the WARP science mode from raw science data. Figure \ref{flow} shows the flow of the data reduction of WARP calibration and science modes. Details on the each reduction process will be given in the following subsections.

For ThAr lamps, the WARP has reference files, where information on the typical positions of apertures and emission lines are recorded. The reference files are prepared for the three observational modes. If the aperture and/or emission lines positions of the input dataset are significantly apart from those in the reference files, the automatic search fails. In this case, the positions of apertures and/or emission lines recorded in the reference files must be adjusted based on the measurement with new dataset.

\subsection{Echelle order map}

\begin{figure*}
\includegraphics[width=18cm,clip]{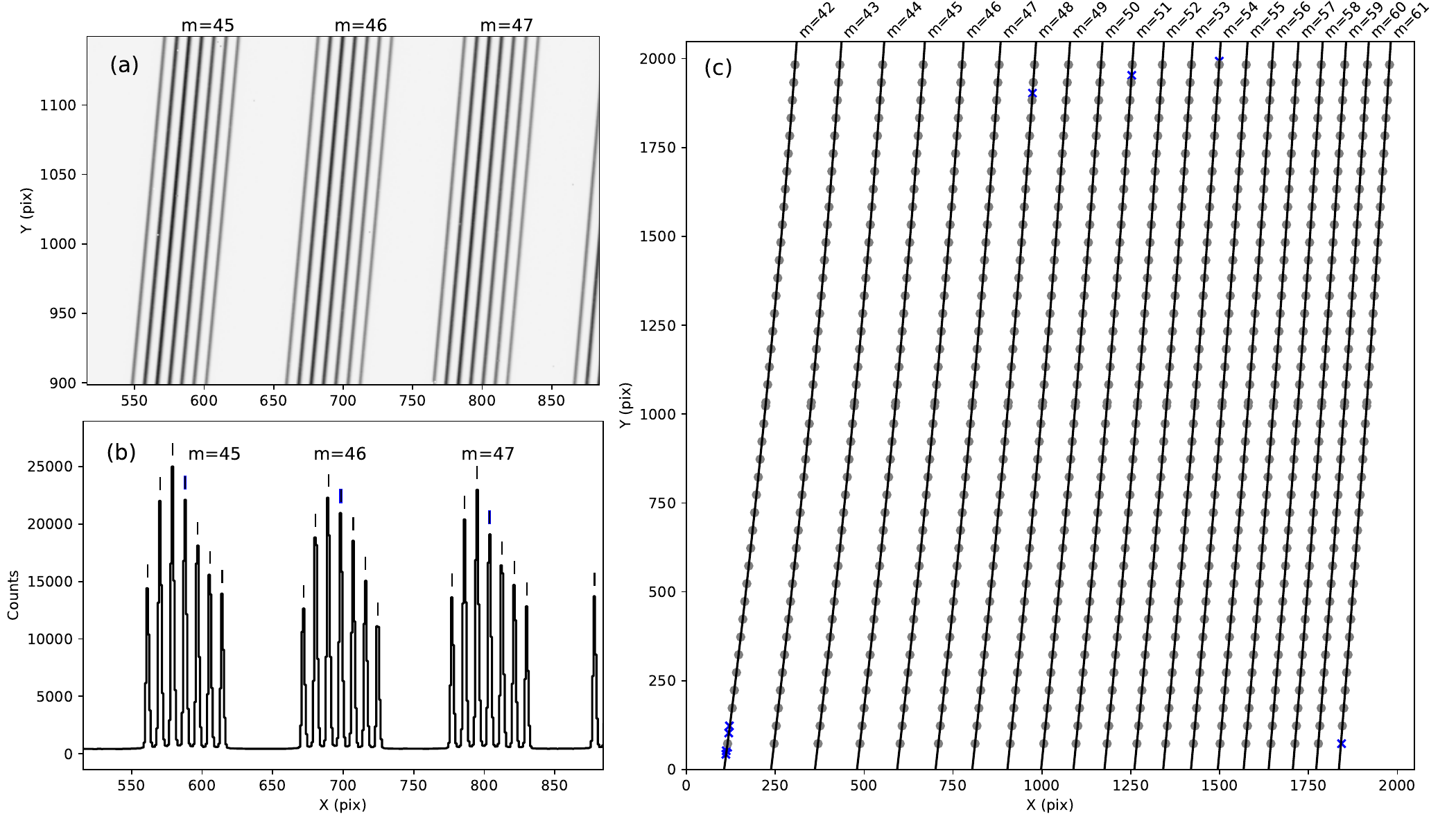}
\caption{Positions of echellogram orders traced by the flat image of multi-pinholes. Panel (a): Data of the flat image of multi-pinholes. The horizontal axis is shared with image (b). Panel (b): Cross-cut plot of the data at $Y=1024$. The peaks detected with the pipeline routine are marked with lines. The central peak of each order is marked with a thick blue line. Panel (c): Positions of the central peaks of each order in the whole array. Data points are fitted with the polynomial functions shown by the lines. Gray circles show the data points used in the fitting while the blue crosses show the data points excluded due to deviation from the fitting function. Only one fifth of the fitted data points are shown in the figure for visibility.}
\label{multiholetrace}
\end{figure*}

To search for the object signal and extract the spectra from the scientific data, both the position and size of the strip of each echelle order in the array are necessary. We use the flat image of the multi-pinholes to trace and identify the echelle order in the array. Figure \ref{multiholetrace} shows the flat image of multi-pinholes. The flat-lamp light, which passes through the seven holes set at the position of the slit, is dispersed by the grating and cross-dispersed by the VPH grating, forming seven continuous lines per echelle order in the image (Figure \ref{multiholetrace}a). Using the lines, the echelle orders are identified and traced.

First, cross-cut data are produced by averaging the data of the five pixel lines around the center of the array. Peaks are detected from the cross-cut data using the scipy routine, {\tt scipy.signal.argrelmax}, which finds the relative maxima from the input data. Then, the gravity centers of the detected peaks are calculated. By comparing the center positions of the detected peaks with those read from reference file, echelle orders will be identified for the peaks detected from input file (Figure \ref{multiholetrace}b). Although there can be a large offset between the measured peak centers in the input file and those recorded in the reference due to the engineering maintenance, echelle orders can be identified by searching for the shift value matching with the peak centers of the input file and the reference file. The shift value is searched within the range of $\pm 30$ pixels.

Once the peaks of multi-pinholes are identified at the center of the array, the curved lines of all echelle orders are traced along the dispersion direction. Although there are seven lines per echelle order, we trace only the signal of the central hole for the simplicity of the routine. Because the seven lines of each echelle order run almost parallel and the widths of the strip of echelle orders are almost constant, it is not necessary to trace all lines. We estimate the uncertainty of peak positions caused by this simplification as $<0.1$ pixel, which is only about 1\% of the typical aperture width of the point source in the previous WINERED observations. The peak positions of the central hole signal are measured by 10 pixel interval in the $Y$ direction and fitted with the 5th-order Legendre polynomial (Figure \ref{multiholetrace}c). The points apart from the fitted line at 5$\sigma$ level are excluded from the fitting. The coefficients of the polynomial are recorded in the ascii file and will be used in the following procedures in the calibration mode and in the reduction of scientific frames. The widths of the strip of echelle orders are set as a constant once determined from the flat-on data.

\subsection{Bad pixel mask}

The routine of WARP makes the bad pixel mask from on- and off-flat images. Because bad pixels can increase by the degradation of the array caused by the thermal cycle for example, the bad pixel mask must be updated regularly.

The pixels, which have peculiarly high or low counts in off-flat images, are identified as bad pixels. In advance of the identification, the median filtered image with the window size of the 15 pixels is subtracted from the original off flat image to highlight peculiar pixels by removing the bias variation pattern.
Then, the pixels with higher or lower values than the thresholds, are picked up as bad pixels. The threshold is set as the 5$\sigma$ from the average of the image. The standard deviation is calculated after removing the pixels with $>$100 counts, which are apparently bad pixels.

Next, the WARP identifies bad pixels from the on-flat image. The on-flat image is fitted by the Legendre polynomial of order 15 along with the dispersion direction. Then, the image is normalized by the function. The pixels with $< 0.5$ or $> 1.5$ counts for the normalized on-flat-fielding image are defined as bad pixels, and their coordinates in the array are recorded. The threshold is empirically defined from the histogram of the normalized on-flat image. Finally, the masks made from the on- and off-flat images are combined to make the final bad pixel mask. The mask will be used for the interpolation of pixel values in the following procedures.

\subsection{Flat-fielding image}

The flat-fielding image is necessary for the correction of pixel-to-pixel variability of the sensitivity. The flat-fielding image is made from the data obtained by observing the flat lamp using the same observing mode. 
First, the counts of bad pixels on the combined on- and off-flat images are replaced with linearly interpolated counts of surrounding normal pixels. Then, the cleaned off-flat image is subtracted from the cleaned on-flat image to remove the bias and dark current. To remove the scattered light in the on-flat-fielding image, the counts of inter-order regions are fitted with a two-dimensional polynomial function and subtracted using the {\tt noao.imred.echelle.apscatter} task of {\tt PyRAF}.  Finally, the flat image is divided by a constant value for the normalization. The resultant image is used in the flat-fielding procedure.

\subsection{Echellogram distortion map}

\begin{figure*}
\includegraphics[width=16cm,clip]{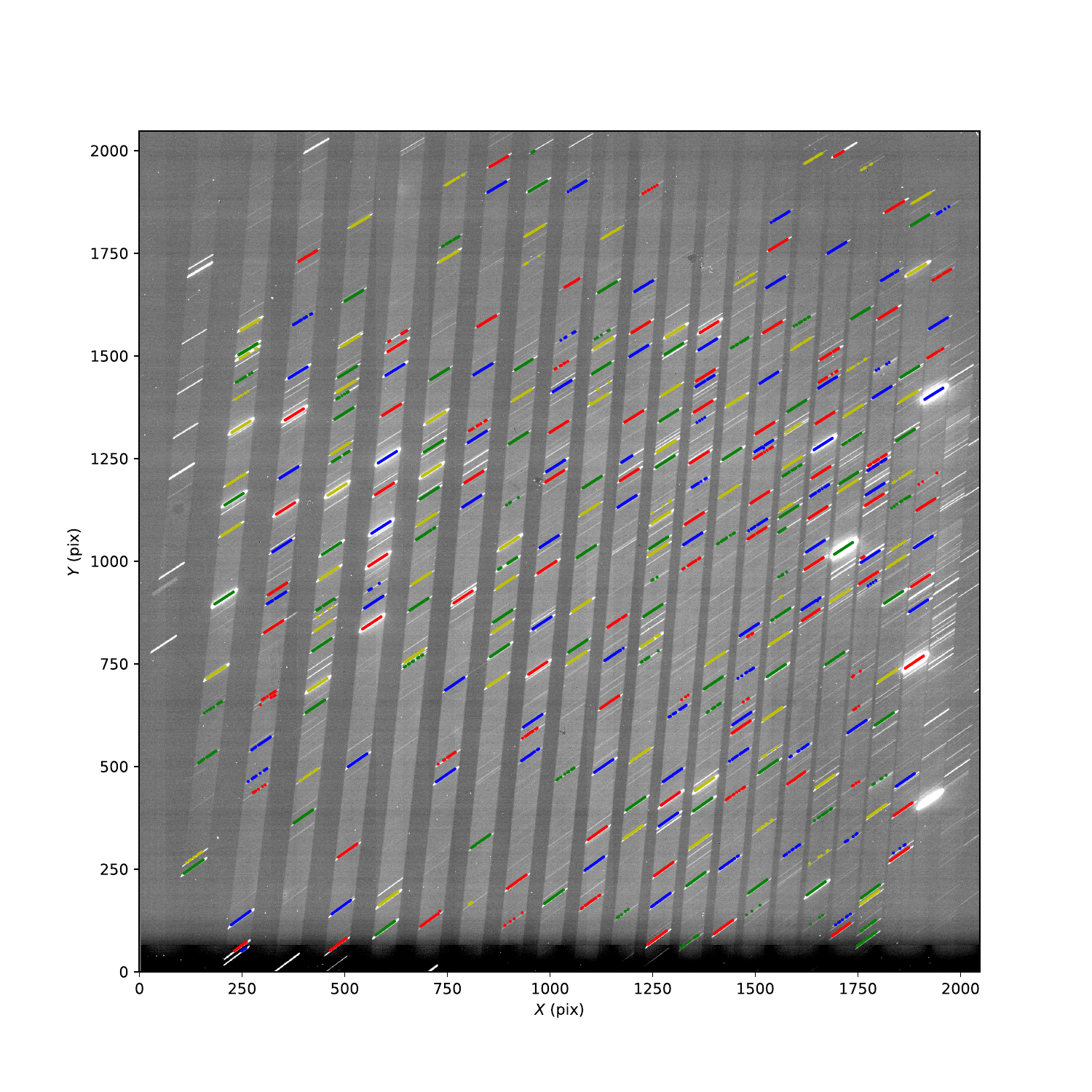}
\caption{Comparison ThAr lamp data and positions of the detected line peaks. The detected line peaks are plotted. The peaks  forming a line are grouped, and the angle of the emission line is measured by fitting a linear function to the peak positions in each group. To make it easier to distinguish the line groups, the peak positions of each group is plotted in alternating colors.}
\label{compmap}
\end{figure*}

\begin{figure*}
\includegraphics[width=16cm,clip]{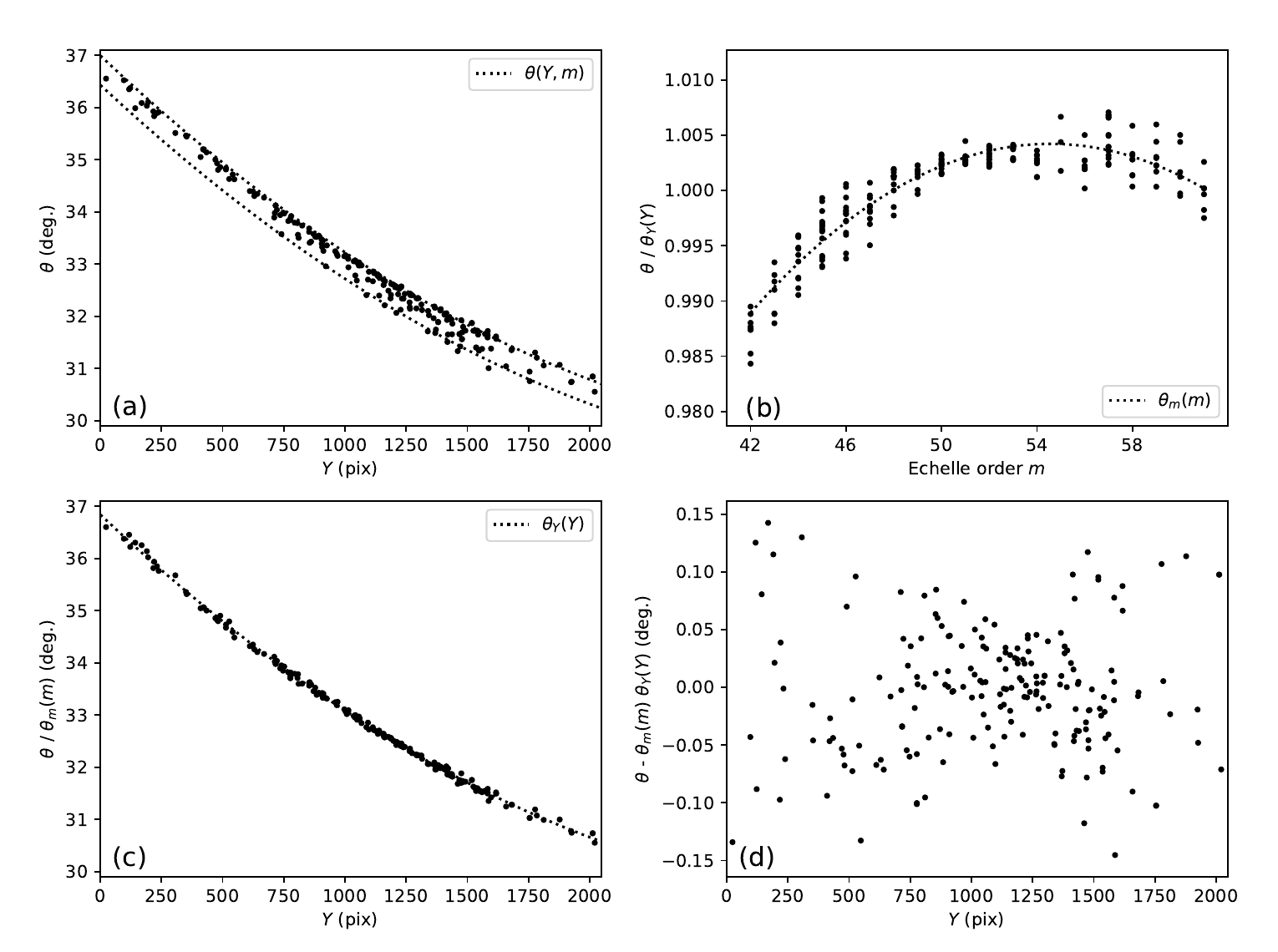}
\caption{Angles of emission lines in the comparison data obtained with the WIDE mode. Panel (a): Angles of emission lines. Dotted lines show the range of the fitted function $\theta (Y, m)$ to the data points (see text for details of the fitting function). Panel (b): Dependency of the angles on echelle order $m$. The angles divided by $\theta _Y (Y)$ are plotted. The dotted line shows the function $\theta _m (m)$ fitted to the data points. Panel (c): Dependency of the angles on $Y$. Angles divided by $\theta _m(m)$ is plotted. The dotted line shows the function $\theta_Y (Y)$ fitted to the data points. Panel (d): Residual of the fitting. }
\label{angle}
\end{figure*}

\begin{figure*}
\includegraphics[width=18cm,clip]{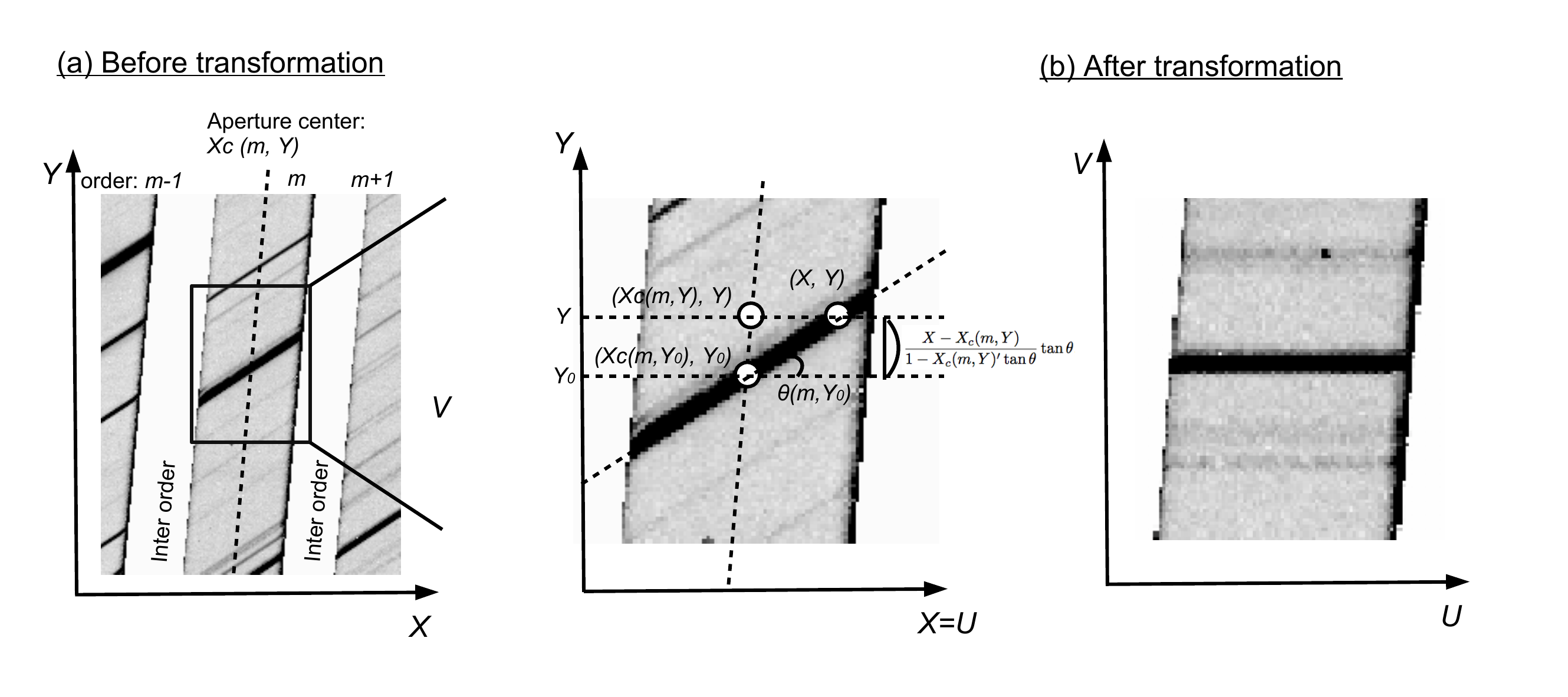}
\caption{Schematic image of the transformation. Panel (a): Comparison data image before the transformation with the geometric descriptions of the equations (2)-(4). Panel (b): Transformed comparison data image.}
\label{transformation}
\end{figure*}

Next, WARP makes the echellogram distortion map, which is used for the distortion correction of the slit images in the raw data. As described in Section 2, the spatial direction of the spectrum is significantly tilted from the axes of the array coordinates. The angles from the $X$ axis are about 30--36 and 42--58 degrees for data obtained by the WIDE and HIRES modes, respectively. The angles depend on both of the array coordinates $(X, Y)$.
The accuracy of the map is essential for the high-dispersion spectroscopic data like WINERED data, as a slight warp in the slit image can cause degradation of the spectral resolution and wavelength accuracy. Using the distortion map, we resample the image to align the spatial direction to the $X$ direction.

The distortion map is made using ThAr emission lines, which are distributed all over the array.
In WINERED data, we assume that the images of emission lines on the array are straight lines in all observational modes to simplify the formulation, as the deviation from the straight line is only about 0.1 pix at most in the WIDE mode. 
Therefore, the shape of the emission line images can be characterized only by the angle from the $X$ axis, $\theta$, which is not constant, but varies with the positions of the emission line images on the array. To obtain the echellogram distortion map, the dependences on the array coordinates must be determined accurately by measuring the angles and center positions of emission line images in the comparison data.

In preparation of the angle measurement, we extract the one-dimensional spectra from the comparison data using the trace function of the echelle orders. The aperture width is set as narrow as 1 pix. The aperture centers in the spatial direction are set as $3i$ pixels ($i=-7, -6, \cdots, 0, \cdots, 6, 7$) offset from the center of the echelle order regions. In total, 15 spectra of comparison lines are extracted from each echelle order.

From the extracted spectra, the peaks are detected by {\tt scipy.signal.find\_peaks\_cwt}, which can discern peaks in a data sequence. Then, the gravity centers of the detected peaks are calculated. The $X$ coordinate corresponding to the peak centers can be calculated from the trace function of each echelle order. From the $(X, Y)$ coordinates of the peaks of emission lines detected from the extracted 15 consecutive spectra, the peaks arising from the same emission lines from consecutive spectra are grouped. Figure \ref{compmap} shows the detected peak positions plotted over the comparison image. The emission lines detected in more than three consecutive spectra are used for the measurement of the angles. The angles are measured by fitting a linear function to the $(X,Y)$ coordinates of each group. Consequently, the set of the parameters characterizing the positions and angles of emission lines, $\theta, X_c, Y_c$, and $m$, can be obtained, where $(X_c, Y_c)$ is the peak positions of the emission lines at the center of the slit.

Figure \ref{angle} shows the relation between the angles and the center positions of the emission lines measured with the WIDE-mode comparison data. 
We expressed the angles of the monochromatic slit image $\theta$ as 
$\theta (m,Y_c) = \theta_m(m) \theta_Y(Y_c),$
where $\theta_m (m)$ and $\theta_Y(Y_c)$ are the quadruple functions of $m$ and $Y_c$, respectively. 
Figure \ref{angle}a shows the angles of emission lines plotted as a function of $Y_c$. Figure \ref{angle}b shows the angles divided by the quadratic function of $Y_c$, $\theta (Y_c)$, fitted to the data points in Figure \ref{angle}a. We observe that $\theta / \theta (Y_c)$ depends on the echelle order $m$. Figure \ref{angle}c shows the angles $\theta$ divided by $\theta_m(m)$ fitted to the data points in Figure \ref{angle}b. The dispersion observed in Figure \ref{angle}c is significantly smaller than that in Figure \ref{angle}a. Figure \ref{angle}d shows the residuals of the fitting. The standard deviation of the residual is about 0.05 deg., which is small enough for the following reduction processes.

The angle function $\theta (m,Y_c)$ is used for the transformation of data to align the slit images parallel to the $X$ direction. Figure \ref{transformation} shows the schematic image of the transformation. The transformation of the coordinates can be expressed by following equations (Figure \ref{transformation}a),
\begin{align}
U &= X, \\
V &= Y - (X - X_c(m,Y_c)) \tan \theta (m, Y_c),\\
  &= Y - \frac{X- X_c(m,Y)}{1 - X_c' (m, Y) \tan \theta (m, Y_c)} \tan \theta (m, Y_c).
\end{align}
In constructing these equations, we added a requirement that the $(X, Y)$ and $(U, V)$ at the slit center match before and after the transformation. We use the IRAF task {\tt noao.twodspec.transform} for this image transformation. The above equations are converted to the form of orthogonal polynomials that can be read from the transformation task. The coefficients of the polynomials are recorded in ascii files. Figure \ref{transformation}b shows the comparison data before and after the transformation. The emission lines align with the horizontal axis of the image after the transformation, which simplifies the spectrum extraction to the integration along the horizontal axis.

\subsection{Dispersion solution}

\begin{figure*}
\includegraphics[width=18cm,clip]{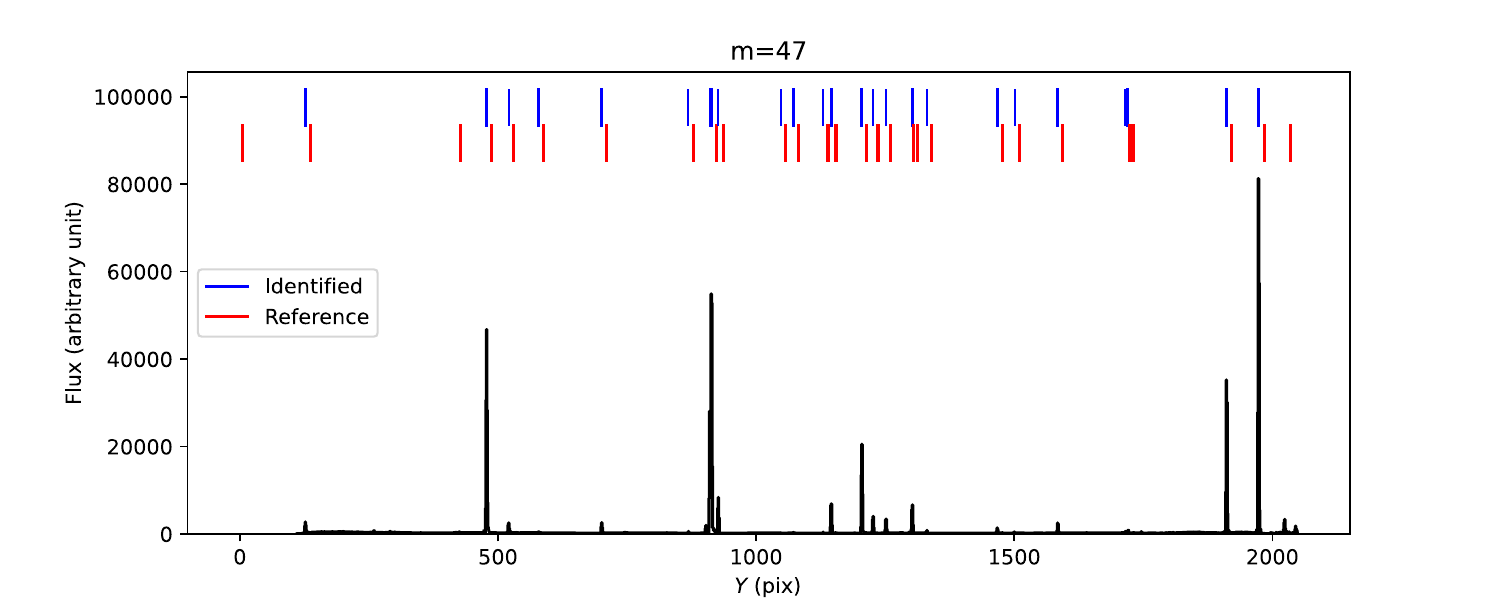}
\caption{Emission spectrum of comparison data ($m=47$ of WIDE mode). Reference line positions are plotted with red lines, and the identified lines in the spectrum are plotted with blue lines. The strong lines used for calculating the offset between the identified line positions and the reference line positions are plotted with thick blue lines. The lines identified after determining the offset are plotted with thin blue lines.}
\label{compID}
\end{figure*}

To obtain the dispersion solution relating pixel positions in the array to wavelengths, the routine identifies the emission lines using the peak positions and the wavelengths of comparison emission lines recorded in the reference file. The list of Th and Ar lines is adopted from \citet{ker08}. Typically about 500, 400, and 270 lines are used for WIDE, HIRES-J, and HIRES-Y modes, respectively. The peak positions in the reference file are the values measured from the comparison spectrum obtained in the specific observational run. 

First, the one-dimensional comparison spectra are extracted by integrating the fluxes along the slit direction from the transformed ThAr lamp image. The routine detects the peaks in the comparison spectrum using  {\tt scipy.signal.find\_peaks\_cwt}. Then, the center positions of the detected lines are compared with those from the reference file, and the detected lines are identified. 
In the identification, the shift of the comparison spectrum in the wavelength direction are considered. Figure \ref{compID} shows the example of the comparison emission spectrum and the identification of the emission lines.
A two-dimensional polynomial function (dispersion solution), $\lambda = \lambda(Y’, m)$ is fitted to the set of peak positions, echelle orders, and the wavelengths of identified emission lines. We use {\tt noao.imred.echelle.ecidentify} of PyRAF for the fitting procedure. The simultaneous determination of the dispersion solution for all of the orders realizes high and uniform wavelength accuracy in the entire wavelength range even for orders with a small number of Th and Ar lines. The final wavelength accuracy is about 1/20 pixel, which is sufficient for our scientific requirements.

\section{Science mode}

The aim of the science mode of WARP is to automatically extract the wavelength-calibrated spectrum from raw data. Using the calibration parameters and data produced by the calibration mode, the following reduction procedures are conducted (Figure \ref{flow}):
\begin{itemize}
\item Sky subtraction
\item Cosmic ray masking
\item Scattered light subtraction
\item Flat fielding
\item Interpolation of bad and cosmic-ray pixels
\item Transformation of spectrum image
\item Spectrum extraction
\item Shift correction by cross-correlation 
\item Wavelength calibration
\item Continuum normalization
\item Combination of all frames
\end{itemize}
Some procedures can be skipped. For example, if the user aims for a quick look into the data during observations and wants to save processing time, it is recommended that some time-consuming procedures not necessary for the quick look are skipped. 

WARP does not conduct the non-linearity correction because WINERED has been operated in the linear region ($<$70 k e$^-$), in which the linearity within $\pm$1\% is guaranteed. Also, WARP does not merge the orders because its necessity and the required procedures depend on the science cases. After the reduction is finished, the information, input parameters and output parameters, such as the number of cosmic ray pixels, the shifts between multiple frames, and the aperture positions and ranges, are summarized in the reduction report pdf file. When the multiple frames are reduced, the signal-to-noise ratio of the combined one-dimensional spectrum is measured by comparing the successive spectra. Because WINERED data suffer from the effect of persistence, which is known to have a different gain value from the accumulated signal \citep{tul19}, it is difficult to estimate the variance of each pixel considering the persistence contamination. Therefore, we could not provide the detailed error information for the extracted spectra yet.

To reduce raw frames with the science mode, an ascii file that lists the filenames of the frames must be prepared. In NIR spectroscopic observations, we obtain the background frames corresponding to the object frames by nodding the telescope in the WINERED observations (see Section 2). Therefore, the object frames and corresponding background frames must be written in the frame list. The frame list can be generated automatically using the observational information in WODB for the dataset taken with the usual sequence, such as ``ABBA'', and ``OSO''. The frame list is passed to WARP as a positional argument in the command line. The paths to the directories, where the raw and calibration data are stored, must be specified with the optional arguments with {\tt -r} ({\tt --rawdatapath}) and {\tt -c} ({\tt --calibpath}), respectively. If these arguments are not set when running the pipeline, both paths are assumed to be current directory by default. Furthermore, the directory path, which is newly made to store the reduced data, must be set as the optional argument with {\tt -d} ({\tt --destpath}). Some other optional arguments can be set additionally. 
For example, {\tt -q} ({\tt --query}) and {\tt -p} ({\tt --parameter}) are the option keywords for setting pipeline parameters. By adding {\tt -s} ({\tt --save}) keyword when running the pipeline, the intermediate files that are created during each process of the pipeline are not deleted. After completing the data reduction processes, the output files are stored in the structured directories and the pipeline report pdf file, which stores the results of the pipeline reduction and information of the reduced data. The detailed processes of each reduction step are described in the remainder of this section.

\subsection{Sky subtraction}

The object frame and corresponding background frame are listed in the inputted frame list. As a first step of the data reduction, the background frame is subtracted from the object frame to remove the background emission, dark current, bias, and stray light, which are sources of systematic noise. For images obtained with ABBA sequence, the frame at the B (or A) position is subtracted from that at the A (or B) position. In the case of the data obtained with OSO sequence, the frame at the S position is subtracted from the O frame. The background spectrum can be subtracted during extraction of the spectra when the background region in a slit is inputted when running the pipeline (see Section 5.6). With the additional subtraction of a background spectrum, the residual of OH emission lines can be subtracted well.

\subsection{Cosmic ray mask}

\begin{figure*}
\includegraphics[width=18cm,clip]{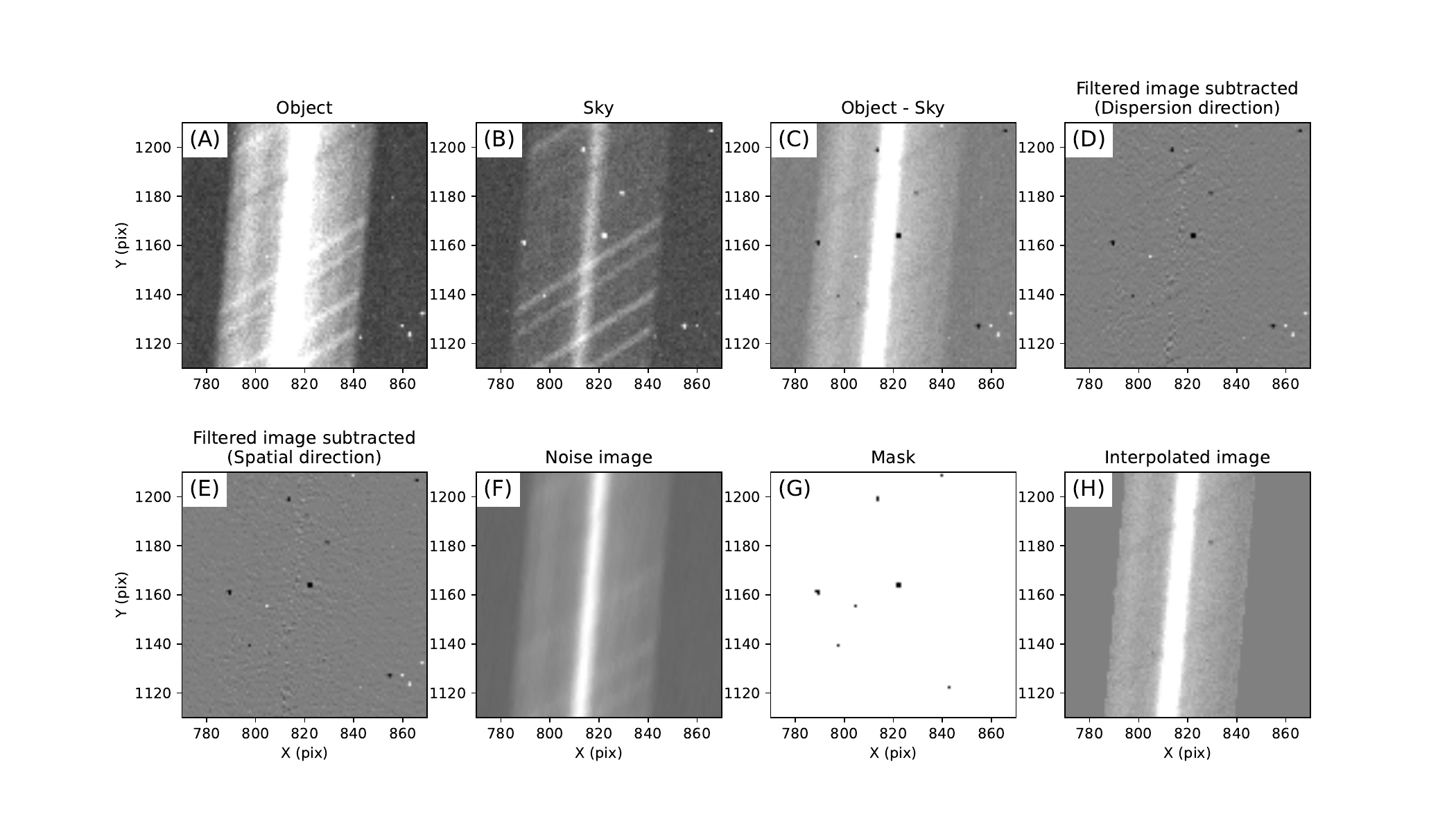}
\caption{Images explaining the cosmic ray detection algorithm. Panels (A) and (B) show the raw data of the object and sky frames, respectively. The slightly tilted line running in the center of the panel (A) shows the spectrum of the object. The fainter line in the same area in the sky frame shows the persistence of the object frame. Some weak sky lines are observed in the both frames. Panel (C) shows the difference of the object and sky frames (A$-$B). Panel (D) shows the (C) image after the subtraction of the (C) image filtered in the direction of the $Y$ axis, and Panel (E) shows the (D) image after the subtraction of the (D) image filtered in the $X+Y$ direction. Panel (F) shows the noise image constructed from the raw data (A) and (B). Panel (G) shows the mask of the cosmic ray detected by the comparison between the images (E) and (F) (see main text for detail). Panel (H) shows the interpolated (object $-$ sky) frame using the mask.}
\label{hotpixmask}
\end{figure*}

The pixels hit by cosmic rays during exposure show a higher signal than the ambient pixels. The cosmic-ray pixels are detected from the (object - sky) frame, and are recorded in the mask file, which will be used in the interpolation of the cosmic-ray pixels (see Section 5.4). Figure \ref{hotpixmask} shows how the pipeline pick up the cosmic-ray pixels. For the detection of the cosmic-ray pixels, the (object - sky) frame is smoothed by the median filter with five pixels along the Y direction, which is almost the dispersion direction. The smoothed image is subtracted from the original (object - sky) frame. Second, the obtained image is smoothed again by the median with five pixel along the 45 degree direction to the X and Y axes, which is almost the monochromatic slit image direction, and the obtained image is subtracted similarly as done for the Y direction smoothed image. Panel (E) of Figure \ref{hotpixmask} shows the residual of the (object - sky) frame after the two-fold subtraction of the median-filtered images. The signal from the object and sky emission are almost removed by the filtering process, and the pixels hit by the cosmic ray are exaggerated in the image. 

The cosmic-ray mask is created by comparing the residual image with the noise image of the (object - sky) frame, which is calculated as follows:
\begin{equation}
\begin{split}
 &N (X, Y) = \\ &g^{-1} \sqrt{g(O(X, Y) + S(X, Y)) + r^2 / NDR_{\text{obj}} + r^2 / NDR_{\text{sky}}}, 
\end{split}
\end{equation}
where $N$ is the noise image, $g$ is the gain (e$^-$/ADU) of the detector, $O$ and $S$ are the counts of the raw data of the object and sky frames, respectively, $r$ is the readout noise, and NDR$_\text{obj}$ and NDR$_\text{sky}$ are the number of the non-destructive readout of the object and sky frames, respectively. The gain, $g$, and readout noise, $r$, are measured as 2.27 e$^-$/ADU and 19.2 e$^-$, respectively \citep{ike22}. As a result of the tests with actual spectroscopic data, the apparent cosmic-ray pixels can be efficiently picked up by recording the pixels whose counts in the residual image are ten times larger than the counts in the noise image. However, in some cases of the bright objects, the normal pixels that have large counts in the residual image, such as the peaks of the sharp point spread functions (PSFs), and the peaks of the narrow emission lines, are erroneously recorded in the cosmic ray mask. To avoid such false detection, a factor to adjust the threshold of cosmic-ray pixels is introduced based on the tests with real spectroscopic data. The threshold of the cosmic rays is multiplied by the factor. The factor is defined as 1 in the array region, where the standard deviation of the residual image is smaller than the noise image. When the standard deviation of the residual image is larger than the noise image, the factor is calculated based on the ratio of the standard deviation to the noise image. Because the factor is set to be larger than 1 in the region with the large residual of the sharp features, the false detection can be suppressed.

After the cosmic ray mask is created, the distribution of the cosmic-ray pixels in the slit direction is calculated for the check of the false detection. If cosmic-ray pixels are concentrated around the PSF peak, this shows that numerous normal pixels around the PSF peak are erroneously identified as the cosmic-ray pixels. When the density of cosmic-ray pixels around the PSF peak exceeds that of the other region by a factor of 1.5, the detection of the cosmic-ray pixels is iterated by setting a larger threshold value in the whole region. The threshold can be fixed by setting pipeline parameters.

\subsection{Scattered light subtraction}

The scattered light due to the random-pitch error and surface roughness of the echelle grating overlaps on the spectrum in the WINERED data. The scattered light is smoothly distributed in the array, and causes the erroneous offset in the final spectrum. The distribution and intensities of the scattered light can be estimated in the inter-order regions of the (object - sky) frame. We use an IRAF task {\tt noao.imred.echelle.apscatter} for the subtraction. To mask the intra-order regions, the trace function made in Section 4.1 is used. We select Legendre polynomials as a fitting function type. The orders of the polynomials are 2nd and 4th in $X$ and $Y$ directions, respectively.

\subsection{Flat fielding and bad pixel interpolation}

The (object - sky) frame and sky frame are divided by the normalized flat fielding image (Section 4.3). Subsequently, the bad pixels and the pixels hit by cosmic rays, both of which are recorded in the mask made in Section 4.2, are linearly interpolated with the counts of the surrounding normal pixels. The IRAF task {\tt fixpix} is used for the interpolation.

\subsection{Transformation of echelle spectra}

The resultant image is transformed to align the monochromatic slit images to the direction of the $X$ axis. We use the {\tt IRAF} task of {\tt transform} in the {\tt noao.twodspec.longslit} package for the transformation. The echellogram distortion map is produced from the ThAr lamp image (Section 4.4). Because this distortion map is separately prepared for each echelle order, the transformation is conducted independently for each echelle order. To reduce the computational time in the transformation procedure, the image size is cut out to rectangles, as it include the whole aperture region of each echelle order, just before the transformation. 
As for the HIRES data, the pixel scale of the $Y$ direction after the transformation is set as the half of the original pixel size to prevent the degradation of the spectral resolution. 

\subsection{Extraction of 1d and 2d spectra}

The center position and width of the aperture must be determined for the extraction of the spectrum from the transformed image. If the aperture range is manually input when running the pipeline, the one-dimensional spectra are extracted from each transformed image by integrating the inputted range in the slit direction. In the case of an extended source or objects that require special treatment (e.g., multiple spectra are observed in the two-dimensional spectra due to multiple stars), the aperture range must be set manually. 

If there is no input of the aperture range, the peak is automatically searched from the transformed images. First, the slit profiles of pixel rows in the transformed image are summed to improve the signal-to-noise ratio. The pixel rows with relatively low counts are not used in the summation. Further, the pixel rows, in which the maxima position is apart from the average of the maximum positions in the slit, are not used in the summation. Because the transformed images are not parallel to the $Y$ direction but slightly curved, the trace functions derived from the calibration data (see Section 4.1) are used for the summation. The peak positions are searched from the summed slit profiles and the FWHMs are measured by fitting a Gaussian profile. The aperture width is set as 2 $\times$ FWHM, which is empirically determined for maximizing the signal-to-noise ratios of the extracted one-dimensional spectra.

If the background region in the slit profile is input when running the pipeline, the background spectrum can be subtracted from the object spectrum. If there is a large residual of the sky emission even after the A $-$ B or O $-$ S subtraction (Section 5.1), the additional background subtraction can reduce systematic noise caused by the sky emission. 

The two-dimensional spectrum is also made at the same time of the extraction of the one-dimensional spectrum. The two-dimensional spectrum can be used for the extended objects. The two-dimensional spectra are extracted by resampling the transformed image after interpolating the counts of each pixel row with the cubic spline function. 

The pipeline simply sums the counts in the aperture range when extracting the one-dimensional spectra. To maximize the signal-to-noise of the resultant spectra, we must integrate the counts by the optimal extraction method, which set weights to the pixels considering the estimated noises in the pixels \citep{hor86,mar89,muk90,pis02}. However, there are two major problems for adopting the optimal extraction in the WINERED pipeline. The first problem is the noise correlation caused by the resampling in the transformation procedure. The noise correlation significantly complicates the calculation of the weights. Further, the resampling causes the systematic noise in the transformed images. The resampling noise is not evidently observed in the extracted one-dimensional spectra, but systematic noise also makes optimal extraction difficult. The data reduction of the distorted echelle spectra without resampling was developed by \citet{pis02}. Although we attempt to adopt their data reduction method, the current reduction flow of the pipeline must be changed drastically and the method may need long calculation time. The second problem is the loss of information of the background noise level. To estimate the background noise, we require the blank fields in the slit. However, the long lasting persistence signal makes it difficult to estimate the background level in many cases. Because the slit is not long enough, it is difficult to maintain some regions in the slit without the object signals for a long time during an observation night. Due to these problems, we could not adopt the optimal extraction method in the pipeline at this time.

\subsection{Image shift by cross-correlation}

The raw two-dimensional spectra of WINERED can slightly shift (within about 0.2 pixels) on the infrared array in the dispersion direction during one night. This originates from the bending of the optical bench due to the variation of the ambient temperature. Because this kind of the shift degrades the spectral resolution and causes correction errors for the telluric absorption correction with a standard star, the relative shifts among multiple frames in a dataset are corrected before combining multiple frames.  

The relative pixel shifts between spectra are found to be almost constant. Therefore, we correct the relative shifts by linearly shifting the spectra in pixels. The shifts of the spectra are measured with the cross-correlation method. The frame that has the highest signal is used as a reference. The shifts are measured independently for each echelle order, and the averaged value is used for the shift correction. The shift values that deviate from the averaged value by two sigma are iteratively removed from the average calculation. The uncertainty of the spectrum shift increases when the lines are very weak and/or very few in the spectrum of an echelle order. Because there are numerous telluric absorption lines in the 0.91--1.35$\mu$m range, most echelle orders are available for the shift measurements. Using the measured shift values, the spectra of each frame are shifted and aligned to the spectrum of the highest-count frame. If the standard deviation of the averaged shift values is larger than a threshold due to the low S/N ratio of the spectrum pair, the procedure does not conduct the shift correction. It is also possible to use the arbitrary shift values by setting the pipeline parameters before running the reduction.

\subsection{Wavelength conversion}

The $Y$-axes of all spectra (all frames and all echelle orders) are converted from pixels to wavelength in vacuum using the dispersion solution made from ThAr lamp images (Section 4.5). Subsequently, the low count regions apart from the free spectral ranges are removed. The spectra with the wavelength in standard air ($T_{\text{air}}$ = 15$^\circ$C, $P_\text{air}$ = 101.325 kPa, Relative humidity = 0 \%, and the CO$_2$ content = 450 ppm) are also produced. We adopted the following conversion form \citet{cid96}, 

\begin{equation}
\lambda _{air} = \lambda _{vac} / \left[ 1.0 + \frac{5.792105 \times 10 ^{-2}}{238.0185 - \lambda _{vac}^{-2}} + \frac{1.67917 \times 10^{-3}}{57.362 - \lambda_{vac}^{-2}} \right] ,
\end{equation}
where $\lambda _{air}$ and $\lambda _{vac}$ are the wavelengths of standard air and vacuum in $\mu$m unit, respectively.

\subsection{Normalization and combination}

The spectra normalized with the continuum level are likewise produced automatically. The quality of continuum fitting may not be good for a science use because the same parameters and fitting function are used for all types of objects. Therefore, the obtained spectra must be limited for the purpose of quick look. Spectra with and without normalization procedure are also produced by simple summation of one-dimensional spectra. These procedures are not applied for two-dimensional spectra. Finally, the pipeline software outputs two two-dimensional spectra in the wavelength units in the air and vacuum ($\lambda _{air}$ and $\lambda _{vac}$), two normalized and two not normalized one-dimensional spectra in the units of $\lambda _{air}$ and $\lambda _{vac}$. 

\subsection*{}
\begin{figure*}
\includegraphics[width=18cm,clip]{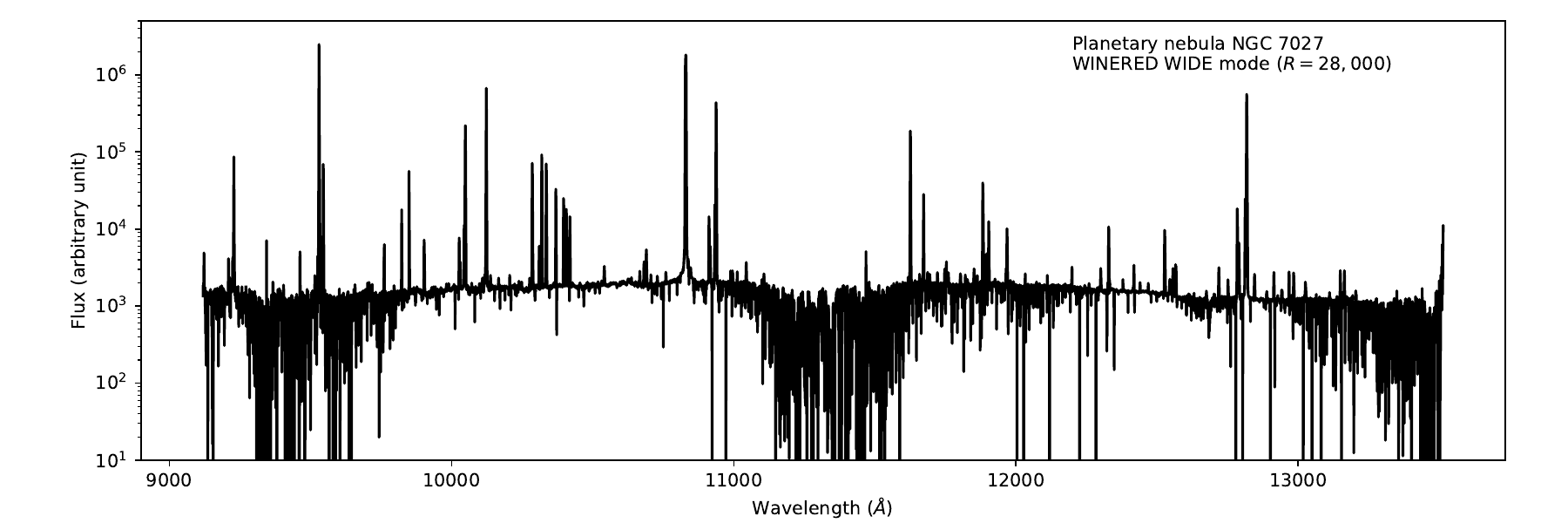}
\includegraphics[width=18cm,clip]{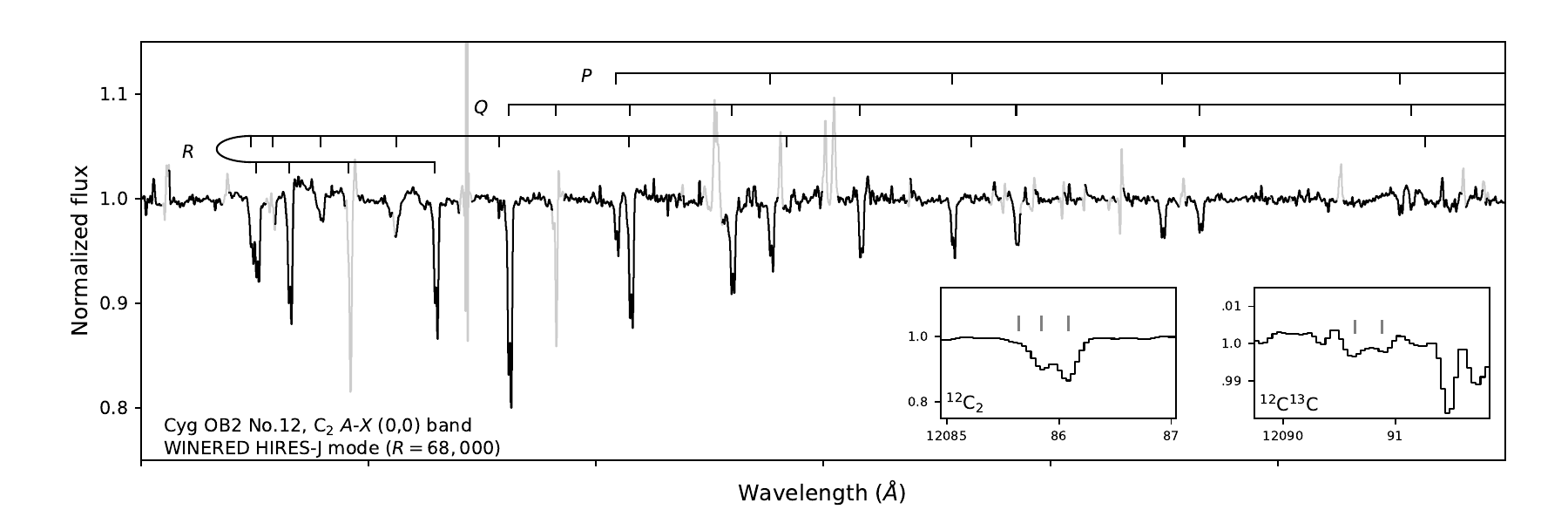}
\caption{Examples of WINERED spectra reduced with WARP. Upper panel: The spectrum of a planetary nebula NGC 7027 obtained with WINERED WIDE mode (Araki 1.3m telescope, $R=28000$, total exposure time of 4,200 sec). Lower panel: The spectrum of a highly reddened early-type star Cyg OB2 No.\,12 obtained with WINERED HIRES-J mode (Araki 1.3m telescope, $R=68000$, total exposure time of 3600 sec). The telluric absorption lines are removed using an A0V star spectrum. The absorption lines of interstellar C$_2$ molecules ($A$--$X$ (0,0) band) are detected. The velocity components of the line-of-sight interstellar clouds are resolved due to the high resolution of WINERED spectrograph for the first time \citep{ham19}.}
\label{finalspec}
\end{figure*}

Figure \ref{finalspec} shows an example of the combined spectra obtained with WINERED WIDE mode reduced with WARP.

\section{Degradation of spectral resolution}
\begin{figure}
\includegraphics[width=9cm,clip]{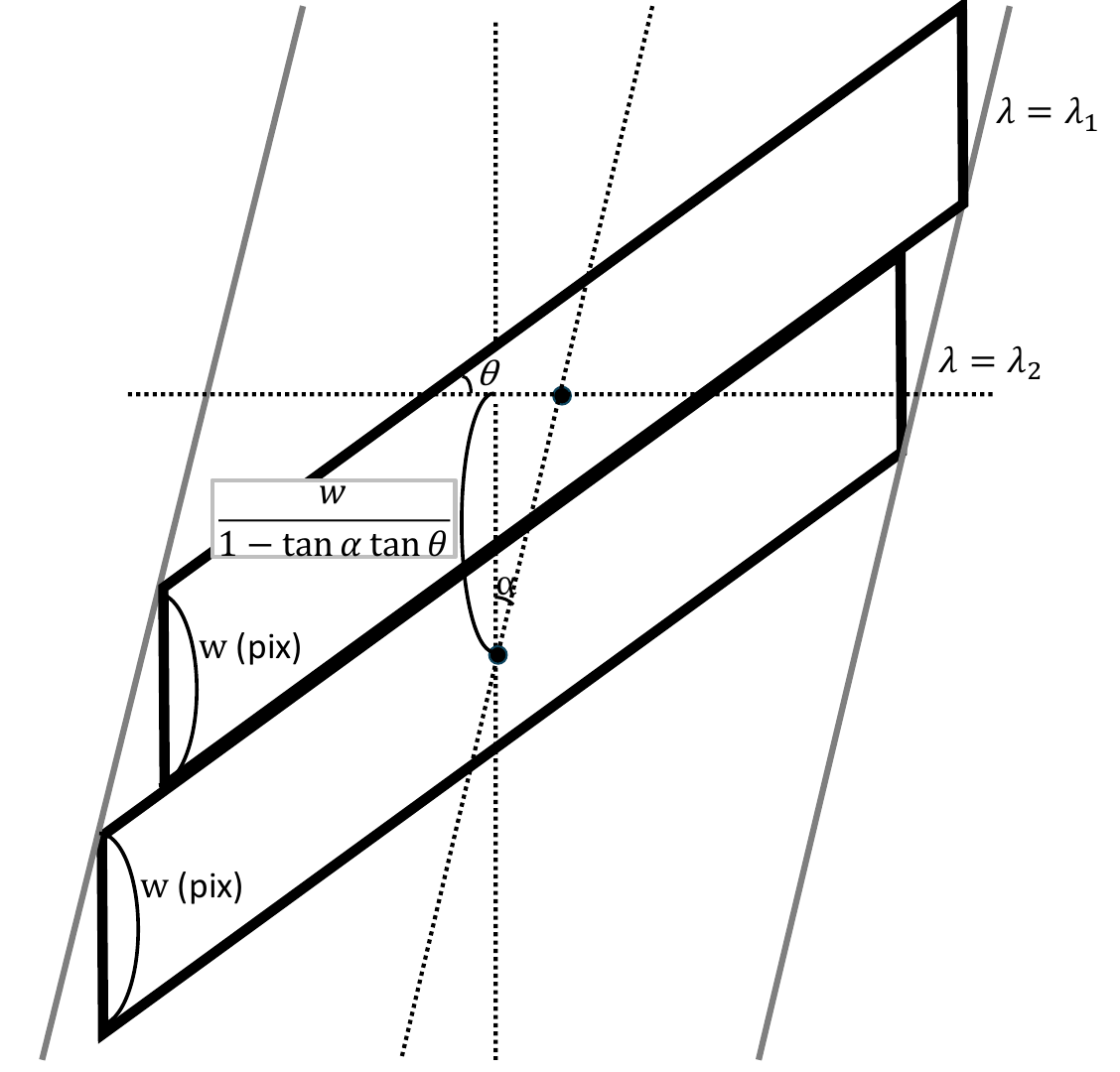}
\caption{Schematic image of distorted lines in WINERED data. }
\label{resolution}
\end{figure}
\begin{figure}
\includegraphics[width=9cm,clip]{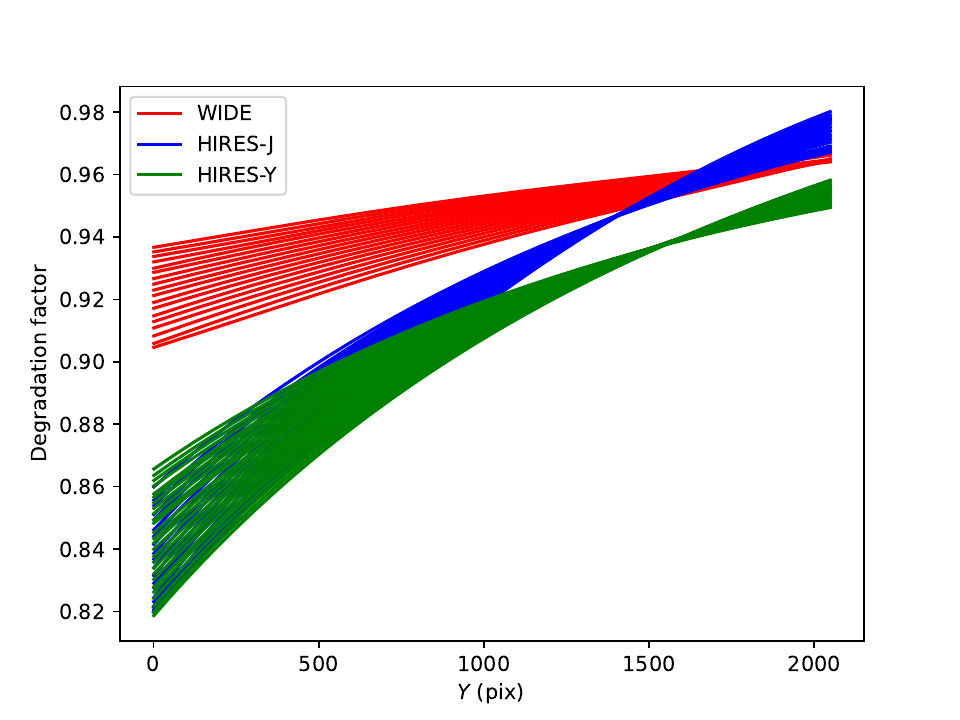}
\caption{The degradation factor of the spectral resolutions. The lines are plotted for each echelle order. The data of WIDE, HIRES-J, and HIRES-Y modes are plotted with red, blue, and green lines, respectively.}
\label{degra}
\end{figure}

\begin{figure}
\includegraphics[width=9cm,clip]{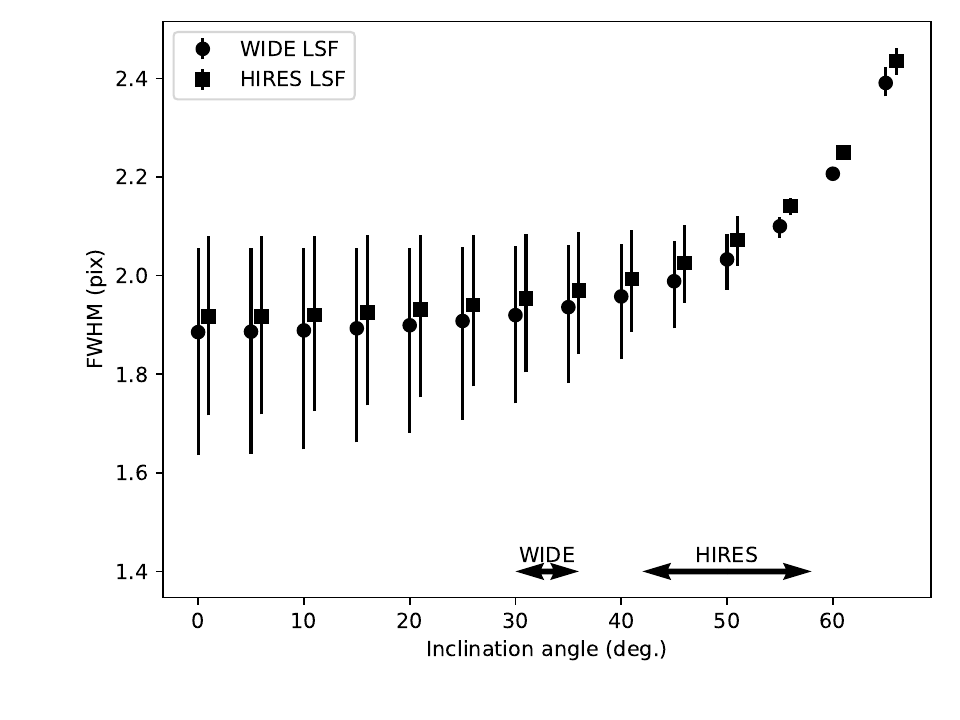}
\caption{FWHMs of pixelized inclined line profiles. The FWHMs of the line profiles of the WIDE and HIRES modes are plotted with the circle and square, respectively. The HIRES points are offset by 1 degree in the horizontal axis direction to avoid overlapping with the WIDE points.}
\label{pixelization}
\end{figure}

\citet{ots16} reported that the spectral resolutions of WINERED HIRES modes are measured as $R=68,000$, which is degraded from the expected spec, $R=80,000$, by a factor of approximately 0.87. In this section, we discuss that the degradation is mainly caused by the under-sampling effect. Note that the possible other causes are proposed in \citet{ike22}: the blurring of the spot diagram of the HIRES mode, the slight tilt of the IR array, and the under sampling effect of seriously squashed and tilted monochromatic slit images by the high blazed echelle grating with a large $\gamma$ angle configuration. They also suggest that the spectral resolution is expected to be recovered to $>68,000$ by the re-alignment of the IR array.

Figure \ref{resolution} shows how the spectral resolutions of WINERED data are degraded from the requirements. The strip of the echelle orders in the WINERED data is slightly inclined from the $Y$-axis direction. The inclination angle is defined as $\alpha$ in Figure \ref{resolution}. The monochromatic slit image is also inclined from the $X$-axis direction. The inclination angle is defined as $\theta$ in Figure \ref{resolution}. We consider the two tangent monochromatic slit images ($\lambda = \lambda _1$ and $\lambda _2$) with the width along the $Y$ direction, $w$. The distance between the $Y$ coordinates of the central points of these lines is expressed as $w/(1 - \tan \alpha \tan \theta)$, which is larger than $w$. Because the dispersion axis is parallel to the $Y$-axis direction, the spectral resolution is degraded by a factor of $1 - \tan \alpha \tan \theta$ with the distortion. The same factor can be derived from the image transformation formulae (1) -- (3). Using the actual data obtained with the WINERED, the degradation factor of the spectral resolutions is plotted for WIDE, HIRES-Y, and HIRES-J modes in Figure \ref{degra}. The spectral resolutions of WIDE and HIRES modes are estimated to be degraded than the expected spec by a factor of approximately 0.94 and 0.91, respectively, at the center of the array.

Furthermore, the pixelization effect can degrade the spectral resolution. We simulated pixelization of 2-D line profiles while varying the tilt angle $\theta$ and measured the FWHM in the dispersion direction by Gaussian fitting. The line profile was made by convolving the line spread function with the window function with 2 pix width, which is the width of the narrowest slit of WINERED. Figure \ref{pixelization} shows the obtained results. The bars show the ranges of the FWHMs for different center positions of the line within a pixel. The FWHMs become larger by a factor of about 1.02 and 1.08 for WIDE and HIRES modes, respectively. As a result, the spectral resolution become degraded by a factor of 0.98 and 0.93 for WIDE and HIRES modes, respectively, due to the pixelization effect.

Combining the spectral degradation of HIRES modes caused by the inclined slit image (0.91), the total degradation factor is roughly estimated as 0.85 ($= 0.91 \times 0.93$), which is consistent with the measured degradation factor, 0.87, from the expected spec. These two effects caused by the highly inclined slit image should be incorporated when the expected spectral resolution is calculated. 

\section{Summary}
We present a data reduction pipeline, WARP, developed for the NIR echelle spectrograph WINERED. The characteristics of the WINERED data are incorporated into WARP. WARP is equipped with two modes: a calibration mode that creates the calibration images and parameters necessary for the reduction of science frames, and a science mode that reduces the raw science frame to a spectrum. Both modes can be executed automatically and contribute to fast production of the quality-controlled spectroscopic data obtained with WINERED. WARP is implemented with Python. 

We discuss possible causes of degradation of spectral resolutions for HIRES modes. The distortion of line shapes in WINERED data reduce spectral resolutions by a factor of 0.94 and 0.91 for WIDE and HIRES modes, respectively, at the array center. The pixelization of inclined line profiles can likewise degrade spectral resolutions especially for HIRES modes by a factor of 0.93. The spectral resolution degradation measured for HIRES mode (0.87) is almost consistent with the total degradation factor of these two effects, 0.85. The spectral resolution degradation caused by the highly inclined line profile must be considered when the expected spec is calculated.

For the instruments that use a non-white-pupil system to maximize the optics throughput, such as WINERED, it is important to keep the tilt angle of the slit images on the detector less than 45 degrees. If the angle exceeds this value, the spectral resolving power decreases significantly due to the distortion and pixelization effects. Additionally, considering the pixelization effect, the spectral sampling of the narrowest slit should be designed to be larger than 2 pixels, the value of WINERED. Also, the large tilt of the slit images caused by the large $\gamma$ angle complicates the data reduction procedures of WARP. Nonetheless, the development of WARP began after the first light observation of WINERED. Therefore, it is crucial to simultaneously develop the data reduction pipeline software alongside the hardware.

\acknowledgments
WINERED was developed by the University of Tokyo and the
Laboratory of Infrared High-resolution spectroscopy (LiH),
Kyoto Sangyo University under the financial supports of
Grants-in-Aid, KAKENHI, from JSPS (Nos. 16684001,
20340042, and 21840052) and the MEXT Supported Program
for the Strategic Research Foundation at Private Universities
(Nos. S081061 and S1411028).
This study is financially supported by Grants-in-Aid, KAKENHI, from JSPS (Nos. 16K17669, and 21K13969).


\begin{thebibliography}{}
\bibitem[Artigau et al.(2014)]{art14} Artigau, {\'E}., Kouach, D., Donati, J.-F., et al.\ 2014, \procspie, 9147, 914715. doi:10.1117/12.2055663
\bibitem[Astropy Collaboration et al.(2013)]{astropy:2013} Astropy Collaboration, Robitaille, T.~P., Tollerud, E.~J., et al.\ 2013, \aap, 558, A33. doi:10.1051/0004-6361/201322068
\bibitem[Astropy Collaboration et al.(2018)]{astropy:2018} Astropy Collaboration, Price-Whelan, A.~M., Sip{\H{o}}cz, B.~M., et al.\ 2018, \aj, 156, 123. doi:10.3847/1538-3881/aabc4f
\bibitem[Ciddor(1996)]{cid96} Ciddor, P.~E.\ 1996, \ao, 35, 1566. doi:10.1364/AO.35.001566
\bibitem[Claudi et al.(2018)]{cla18} Claudi, R., Benatti, S., Carleo, I., et al.\ 2018, \procspie, 10702, 107020Z. doi:10.1117/12.2312555
\bibitem[D'Orazi et al.(2018)]{dor18} D'Orazi, V., Magurno, D., Bono, G., et al.\ 2018, \apjl, 855, L9. doi:10.3847/2041-8213/aab100
\bibitem[Dorn et al.(2023)]{dor23} Dorn, R.~J., Bristow, P., Smoker, J.~V., et al.\ 2023, \aap, 671, A24. doi:10.1051/0004-6361/202245217
\bibitem[Follert et al.(2014)]{fol14} Follert, R., Dorn, R.~J., Oliva, E., et al.\ 2014, \procspie, 9147, 914719. doi:10.1117/12.2054197
\bibitem[Fukue et al.(2021)]{fuk21} Fukue, K., Matsunaga, N., Kondo, S., et al.\ 2021, \apj, 913, 62. doi:10.3847/1538-4357/abf0b1
\bibitem[Hamano et al.(2019)]{ham19} Hamano, S., Kawakita, H., Kobayashi, N., et al.\ 2019, \apj, 881, 143. doi:10.3847/1538-4357/ab2e0f
\bibitem[Hamano et al.(2022)]{ham22} Hamano, S., Kobayashi, N., Kawakita, H., et al.\ 2022, \apjs, 262, 2. doi:10.3847/1538-4365/ac7567
\bibitem[Horne(1986)]{hor86} Horne, K.\ 1986, \pasp, 98, 609. doi:10.1086/131801
\bibitem[Ikeda et al.(2016)]{ike16} Ikeda, Y., Kobayashi, N., Kondo, S., et al.\ 2016, \procspie, 9908, 99085Z 
\bibitem[Ikeda et al.(2022)]{ike22} Ikeda, Y., Kondo, S., Otsubo, S., et al.\ 2022, \pasp, 134, 015004. doi:10.1088/1538-3873/ac1c5f
\bibitem[Jian et al.(2020)]{jia20} Jian, M., Taniguchi, D., Matsunaga, N., et al.\ 2020, \mnras, 494, 1724. doi:10.1093/mnras/staa834
\bibitem[Kaeufl et al.(2004)]{kae04} Kaeufl, H.-U., Ballester, P., Biereichel, P., et al.\ 2004, \procspie, 5492, 1218. doi:10.1117/12.551480
\bibitem[Kelson(2003)]{kel03} Kelson, D.~D.\ 2003, \pasp, 115, 688. doi:10.1086/375502
\bibitem[Kerber et al.(2008)]{ker08} Kerber, F., Nave, G., \& Sansonetti, C.~J.\ 2008, \apjs, 178, 374. doi:10.1086/590111
\bibitem[Kondo et al.(2019)]{kon19} Kondo, S., Fukue, K., Matsunaga, N., et al.\ 2019, \apj, 875, 129. doi:10.3847/1538-4357/ab0ec4
\bibitem[Marsh(1989)]{mar89} Marsh, T.~R.\ 1989, \pasp, 101, 1032. doi:10.1086/132570
\bibitem[Matsunaga et al.(2022)]{mat22} Matsunaga, N., Itane, A., Hattori, K., et al.\ 2022, \apj, 925, 10. doi:10.3847/1538-4357/ac3483
\bibitem[Matsunaga et al.(2023)]{mat23} Matsunaga, N., Taniguchi, D., Elgueta, S.~S., et al.\ 2023, \apj, 954, 198. doi:10.3847/1538-4357/aced93
\bibitem[Mizumoto et al.(2018)]{miz18} Mizumoto, M., Kobayashi, N., Hamano, S., et al.\ 2018, \mnras, 481, 793. doi:10.1093/mnras/sty2239
\bibitem[Mizumoto et al.(2023)]{miz23} Mizumoto, M., Sameshima, H., Kobayashi, N., et al.\ 2023, arXiv:2311.13085. doi:10.48550/arXiv.2311.13085
\bibitem[Mukai et al.(1990)]{muk90} Mukai, K., Mason, K.~O., Howell, S.~B., et al.\ 1990, \mnras, 245, 385
\bibitem[Origlia et al.(2014)]{ori14} Origlia, L., Oliva, E., Baffa, C., et al.\ 2014, \procspie, 9147, 91471E. doi:10.1117/12.2054743
\bibitem[Otsubo et al.(2016)]{ots16} Otsubo, S., Ikeda, Y., Kobayashi, N., et al.\ 2016, \procspie, 9908, 990879. doi:10.1117/12.2233845
\bibitem[Piskunov \& Valenti(2002)]{pis02} Piskunov, N.~E. \& Valenti, J.~A.\ 2002, \aap, 385, 1095. doi:10.1051/0004-6361:20020175
\bibitem[Piskunov et al.(2021)]{pis21} Piskunov, N., Wehrhahn, A., \& Marquart, T.\ 2021, \aap, 646, A32. doi:10.1051/0004-6361/202038293
\bibitem[Quirrenbach et al.(2013)]{qui13} Quirrenbach, A., Amado, P.~J., Caballero, J.~A., et al.\ 2013, European Physical Journal Web of Conferences, 47, 05006. doi:10.1051/epjconf/20134705006
\bibitem[Rayner et al.(2012)]{ray12} Rayner, J., Bond, T., Bonnet, M., et al.\ 2012, \procspie, 8446, 84462C. doi:10.1117/12.925511
\bibitem[Rayner et al.(2022)]{ray22} Rayner, J., Tokunaga, A., Jaffe, D., et al.\ 2022, \pasp, 134, 015002. doi:10.1088/1538-3873/ac3cb4
\bibitem[Sameshima et al.(2018)]{sam18} Sameshima, H., Matsunaga, N., Kobayashi, N., et al.\ 2018, \pasp, 130, 074502 
\bibitem[Sameshima et al.(2020)]{sam20} Sameshima, H., Yoshii, Y., Matsunaga, N., et al.\ 2020, \apj, 904, 162. doi:10.3847/1538-4357/abc33b
\bibitem[Shinnaka et al.(2017)]{shi17} Shinnaka, Y., Kawakita, H., Kondo, S., et al.\ 2017, \aj, 154, 45. doi:10.3847/1538-3881/aa7576
\bibitem[Taniguchi et al.(2021)]{tan21} Taniguchi, D., Matsunaga, N., Jian, M., et al.\ 2021, \mnras, 502, 4210. doi:10.1093/mnras/staa3855
\bibitem[Tulloch et al.(2019)]{tul19} Tulloch, S., George, E., \& ESO Detector Systems Group\ 2019, Journal of Astronomical Telescopes, Instruments, and Systems, 5, 036004. doi:10.1117/1.JATIS.5.3.036004
\bibitem[Yasui et al.(2019)]{yas19} Yasui, C., Hamano, S., Fukue, K., et al.\ 2019, \apj, 886, 115. doi:10.3847/1538-4357/ab45ee


\end{thebibliography}
\end{document}